\documentclass[12pt]{article}
\usepackage{epsfig}
\usepackage{amssymb}
\usepackage{cite}

\graphicspath{%
}
\DeclareGraphicsExtensions{.eps,.ps,.eps.gz,.ps.gz}
\newcommand{\unit}[1]{\,{\mathrm{#1}}}

\begin{document}
\thispagestyle{empty}

\begin{titlepage}
\begin{center}
EUROPEAN ORGANIZATION FOR NUCLEAR RESEARCH (CERN)
\end{center}
\begin{flushright}

\bigskip
\bigskip
\bigskip

CERN-EP 2003-025\\
May, 20 2003 
\end{flushright}

\bigskip
\bigskip
\bigskip
\bigskip
\bigskip

\begin{center}
  {\LARGE\textbf{Study of hadronic final states from \\[3mm]
      double tagged \boldmath $\gamma \gamma$ events at LEP}}
\end{center}

\bigskip
\bigskip
\bigskip

\begin{center}
{\large The ALEPH Collaboration$^{*)}$}
\end{center}

\bigskip
\bigskip
\bigskip

\begin{abstract}
\noindent
The interaction of virtual photons is investigated using 
double tagged $\gamma \gamma$ events
with hadronic final states recorded by the ALEPH experiment 
at $\mathrm{e}^+ \mathrm{e}^-$ centre-of-mass energies 
between $188$ and $209\,\mathrm{GeV}$. 
The measured cross section is compared to Monte Carlo models, 
and to next-to-leading-order QCD and BFKL calculations.

\bigskip
\bigskip
\bigskip

\begin{center}
{\textit{Submitted to The European Physical Journal C}}\\
\end{center}

\end{abstract}

\vfill

~~~~~$^*)$ See next pages for the list of authors
\end{titlepage}

\pagestyle{empty}
\newpage
\small
%
%
\newlength{\saveparskip}
\newlength{\savetextheight}
\newlength{\savetopmargin}
\newlength{\savetextwidth}
\newlength{\saveoddsidemargin}
\newlength{\savetopsep}
\setlength{\saveparskip}{\parskip}
\setlength{\savetextheight}{\textheight}
\setlength{\savetopmargin}{\topmargin}
\setlength{\savetextwidth}{\textwidth}
\setlength{\saveoddsidemargin}{\oddsidemargin}
\setlength{\savetopsep}{\topsep}
%
%
\setlength{\parskip}{0.0cm}
\setlength{\textheight}{25.0cm}
\setlength{\topmargin}{-1.5cm}
\setlength{\textwidth}{16 cm}
\setlength{\oddsidemargin}{-0.0cm}
\setlength{\topsep}{1mm}
\pretolerance=10000
\centerline{\large\bf The ALEPH Collaboration}
\footnotesize
\vspace{0.5cm}
{\raggedbottom
\begin{sloppypar}
\samepage\noindent
A.~Heister,
S.~Schael
\nopagebreak
\begin{center}
\parbox{15.5cm}{\sl\samepage
Physikalisches Institut das RWTH-Aachen, D-52056 Aachen, Germany}
\end{center}\end{sloppypar}
\vspace{2mm}
\begin{sloppypar}
\noindent
R.~Barate,
R.~Bruneli\`ere,
I.~De~Bonis,
D.~Decamp,
C.~Goy,
S.~Jezequel,
J.-P.~Lees,
F.~Martin,
E.~Merle,
\mbox{M.-N.~Minard},
B.~Pietrzyk,
B.~Trocm\'e
\nopagebreak
\begin{center}
\parbox{15.5cm}{\sl\samepage
Laboratoire de Physique des Particules (LAPP), IN$^{2}$P$^{3}$-CNRS,
F-74019 Annecy-le-Vieux Cedex, France}
\end{center}\end{sloppypar}
\vspace{2mm}
\begin{sloppypar}
\noindent
S.~Bravo,
M.P.~Casado,
M.~Chmeissani,
J.M.~Crespo,
E.~Fernandez,
M.~Fernandez-Bosman,
Ll.~Garrido,$^{15}$
M.~Martinez,
A.~Pacheco,
H.~Ruiz
\nopagebreak
\begin{center}
\parbox{15.5cm}{\sl\samepage
Institut de F\'{i}sica d'Altes Energies, Universitat Aut\`{o}noma
de Barcelona, E-08193 Bellaterra (Barcelona), Spain$^{7}$}
\end{center}\end{sloppypar}
\vspace{2mm}
\begin{sloppypar}
\noindent
A.~Colaleo,
D.~Creanza,
N.~De~Filippis,
M.~de~Palma,
G.~Iaselli,
G.~Maggi,
M.~Maggi,
S.~Nuzzo,
A.~Ranieri,
G.~Raso,$^{24}$
F.~Ruggieri,
G.~Selvaggi,
L.~Silvestris,
P.~Tempesta,
A.~Tricomi,$^{3}$
G.~Zito
\nopagebreak
\begin{center}
\parbox{15.5cm}{\sl\samepage
Dipartimento di Fisica, INFN Sezione di Bari, I-70126 Bari, Italy}
\end{center}\end{sloppypar}
\vspace{2mm}
\begin{sloppypar}
\noindent
X.~Huang,
J.~Lin,
Q. Ouyang,
T.~Wang,
Y.~Xie,
R.~Xu,
S.~Xue,
J.~Zhang,
L.~Zhang,
W.~Zhao
\nopagebreak
\begin{center}
\parbox{15.5cm}{\sl\samepage
Institute of High Energy Physics, Academia Sinica, Beijing, The People's
Republic of China$^{8}$}
\end{center}\end{sloppypar}
\vspace{2mm}
\begin{sloppypar}
\noindent
D.~Abbaneo,
T.~Barklow,$^{26}$
O.~Buchm\"uller,$^{26}$
M.~Cattaneo,
B.~Clerbaux,$^{23}$
H.~Drevermann,
R.W.~Forty,
M.~Frank,
F.~Gianotti,
J.B.~Hansen,
J.~Harvey,
D.E.~Hutchcroft,$^{30}$,
P.~Janot,
B.~Jost,
M.~Kado,$^{2}$
P.~Mato,
A.~Moutoussi,
F.~Ranjard,
L.~Rolandi,
D.~Schlatter,
G.~Sguazzoni,
W.~Tejessy,
F.~Teubert,
A.~Valassi,
I.~Videau
\nopagebreak
\begin{center}
\parbox{15.5cm}{\sl\samepage
European Laboratory for Particle Physics (CERN), CH-1211 Geneva 23,
Switzerland}
\end{center}\end{sloppypar}
\vspace{2mm}
\begin{sloppypar}
\noindent
F.~Badaud,
S.~Dessagne,
A.~Falvard,$^{20}$
D.~Fayolle,
P.~Gay,
J.~Jousset,
B.~Michel,
S.~Monteil,
D.~Pallin,
J.M.~Pascolo,
P.~Perret
\nopagebreak
\begin{center}
\parbox{15.5cm}{\sl\samepage
Laboratoire de Physique Corpusculaire, Universit\'e Blaise Pascal,
IN$^{2}$P$^{3}$-CNRS, Clermont-Ferrand, F-63177 Aubi\`{e}re, France}
\end{center}\end{sloppypar}
\vspace{2mm}
\begin{sloppypar}
\noindent
J.D.~Hansen,
J.R.~Hansen,
P.H.~Hansen,
A.C.~Kraan,
B.S.~Nilsson
\nopagebreak
\begin{center}
\parbox{15.5cm}{\sl\samepage
Niels Bohr Institute, 2100 Copenhagen, DK-Denmark$^{9}$}
\end{center}\end{sloppypar}
\vspace{2mm}
\begin{sloppypar}
\noindent
A.~Kyriakis,
C.~Markou,
E.~Simopoulou,
A.~Vayaki,
K.~Zachariadou
\nopagebreak
\begin{center}
\parbox{15.5cm}{\sl\samepage
Nuclear Research Center Demokritos (NRCD), GR-15310 Attiki, Greece}
\end{center}\end{sloppypar}
\vspace{2mm}
\begin{sloppypar}
\noindent
A.~Blondel,$^{12}$
\mbox{J.-C.~Brient},
F.~Machefert,
A.~Roug\'{e},
M.~Swynghedauw,
R.~Tanaka
\linebreak
H.~Videau
\nopagebreak
\begin{center}
\parbox{15.5cm}{\sl\samepage
Laoratoire Leprince-Ringuet, Ecole
Polytechnique, IN$^{2}$P$^{3}$-CNRS, \mbox{F-91128} Palaiseau Cedex, France}
\end{center}\end{sloppypar}
\vspace{2mm}
\begin{sloppypar}
\noindent
V.~Ciulli,
E.~Focardi,
G.~Parrini
\nopagebreak
\begin{center}
\parbox{15.5cm}{\sl\samepage
Dipartimento di Fisica, Universit\`a di Firenze, INFN Sezione di Firenze,
I-50125 Firenze, Italy}
\end{center}\end{sloppypar}
\vspace{2mm}
\begin{sloppypar}
\noindent
A.~Antonelli,
M.~Antonelli,
G.~Bencivenni,
F.~Bossi,
G.~Capon,
F.~Cerutti,
V.~Chiarella,
P.~Laurelli,
G.~Mannocchi,$^{5}$
G.P.~Murtas,
L.~Passalacqua
\nopagebreak
\begin{center}
\parbox{15.5cm}{\sl\samepage
Laboratori Nazionali dell'INFN (LNF-INFN), I-00044 Frascati, Italy}
\end{center}\end{sloppypar}
\vspace{2mm}
\begin{sloppypar}
\noindent
J.~Kennedy,
J.G.~Lynch,
P.~Negus,
V.~O'Shea,
A.S.~Thompson
\nopagebreak
\begin{center}
\parbox{15.5cm}{\sl\samepage
Department of Physics and Astronomy, University of Glasgow, Glasgow G12
8QQ,United Kingdom$^{10}$}
\end{center}\end{sloppypar}
\vspace{2mm}
\begin{sloppypar}
\noindent
S.~Wasserbaech
\nopagebreak
\begin{center}
\parbox{15.5cm}{\sl\samepage
Utah Valley State College, Orem, UT 84058, U.S.A.}
\end{center}\end{sloppypar}
\vspace{2mm}
\begin{sloppypar}
\noindent
R.~Cavanaugh,$^{4}$
S.~Dhamotharan,$^{21}$
C.~Geweniger,
P.~Hanke,
V.~Hepp,
E.E.~Kluge,
A.~Putzer,
H.~Stenzel,
K.~Tittel,
M.~Wunsch$^{19}$
\nopagebreak
\begin{center}
\parbox{15.5cm}{\sl\samepage
Kirchhoff-Institut f\"ur Physik, Universit\"at Heidelberg, D-69120
Heidelberg, Germany$^{16}$}
\end{center}\end{sloppypar}
\vspace{2mm}
\begin{sloppypar}
\noindent
R.~Beuselinck,
W.~Cameron,
G.~Davies,
P.J.~Dornan,
M.~Girone,$^{1}$
R.D.~Hill,
N.~Marinelli,
J.~Nowell,
S.A.~Rutherford,
J.K.~Sedgbeer,
J.C.~Thompson,$^{14}$
R.~White
\nopagebreak
\begin{center}
\parbox{15.5cm}{\sl\samepage
Department of Physics, Imperial College, London SW7 2BZ,
United Kingdom$^{10}$}
\end{center}\end{sloppypar}
\vspace{2mm}
\begin{sloppypar}
\noindent
V.M.~Ghete,
P.~Girtler,
E.~Kneringer,
D.~Kuhn,
G.~Rudolph
\nopagebreak
\begin{center}
\parbox{15.5cm}{\sl\samepage
Institut f\"ur Experimentalphysik, Universit\"at Innsbruck, A-6020
Innsbruck, Austria$^{18}$}
\end{center}\end{sloppypar}
\vspace{2mm}
\begin{sloppypar}
\noindent
E.~Bouhova-Thacker,
C.K.~Bowdery,
D.P.~Clarke,
G.~Ellis,
A.J.~Finch,
F.~Foster,
G.~Hughes,
R.W.L.~Jones,
M.R.~Pearson,
N.A.~Robertson,
M.~Smizanska
\nopagebreak
\begin{center}
\parbox{15.5cm}{\sl\samepage
Department of Physics, University of Lancaster, Lancaster LA1 4YB,
United Kingdom$^{10}$}
\end{center}\end{sloppypar}
\vspace{2mm}
\begin{sloppypar}
\noindent
O.~van~der~Aa,
C.~Delaere,$^{28}$
G.Leibenguth,$^{31}$
V.~Lemaitre$^{29}$
\nopagebreak
\begin{center}
\parbox{15.5cm}{\sl\samepage
Institut de Physique Nucl\'eaire, D\'epartement de Physique, Universit\'e Catholique de Louvain, 1348 Louvain-la-Neuve, Belgium}
\end{center}\end{sloppypar}
\vspace{2mm}
\begin{sloppypar}
\noindent
U.~Blumenschein,
F.~H\"olldorfer,
K.~Jakobs,
F.~Kayser,
K.~Kleinknecht,
A.-S.~M\"uller,
B.~Renk,
H.-G.~Sander,
S.~Schmeling,
H.~Wachsmuth,
C.~Zeitnitz,
T.~Ziegler
\nopagebreak
\begin{center}
\parbox{15.5cm}{\sl\samepage
Institut f\"ur Physik, Universit\"at Mainz, D-55099 Mainz, Germany$^{16}$}
\end{center}\end{sloppypar}
\vspace{2mm}
\begin{sloppypar}
\noindent
A.~Bonissent,
P.~Coyle,
C.~Curtil,
A.~Ealet,
D.~Fouchez,
P.~Payre,
A.~Tilquin
\nopagebreak
\begin{center}
\parbox{15.5cm}{\sl\samepage
Centre de Physique des Particules de Marseille, Univ M\'editerran\'ee,
IN$^{2}$P$^{3}$-CNRS, F-13288 Marseille, France}
\end{center}\end{sloppypar}
\vspace{2mm}
\begin{sloppypar}
\noindent
F.~Ragusa
\nopagebreak
\begin{center}
\parbox{15.5cm}{\sl\samepage
Dipartimento di Fisica, Universit\`a di Milano e INFN Sezione di
Milano, I-20133 Milano, Italy.}
\end{center}\end{sloppypar}
\vspace{2mm}
\begin{sloppypar}
\noindent
A.~David,
H.~Dietl,
G.~Ganis,$^{27}$
K.~H\"uttmann,
G.~L\"utjens,
W.~M\"anner,
\mbox{H.-G.~Moser},
R.~Settles,
M.~Villegas,
G.~Wolf
\nopagebreak
\begin{center}
\parbox{15.5cm}{\sl\samepage
Max-Planck-Institut f\"ur Physik, Werner-Heisenberg-Institut,
D-80805 M\"unchen, Germany\footnotemark[16]}
\end{center}\end{sloppypar}
\vspace{2mm}
\begin{sloppypar}
\noindent
J.~Boucrot,
O.~Callot,
M.~Davier,
L.~Duflot,
\mbox{J.-F.~Grivaz},
Ph.~Heusse,
A.~Jacholkowska,$^{6}$
L.~Serin,
\mbox{J.-J.~Veillet}
\nopagebreak
\begin{center}
\parbox{15.5cm}{\sl\samepage
Laboratoire de l'Acc\'el\'erateur Lin\'eaire, Universit\'e de Paris-Sud,
IN$^{2}$P$^{3}$-CNRS, F-91898 Orsay Cedex, France}
\end{center}\end{sloppypar}
\vspace{2mm}
\begin{sloppypar}
\noindent
P.~Azzurri, 
G.~Bagliesi,
T.~Boccali,
L.~Fo\`a,
A.~Giammanco,
A.~Giassi,
F.~Ligabue,
A.~Messineo,
F.~Palla,
G.~Sanguinetti,
A.~Sciab\`a,
P.~Spagnolo
R.~Tenchini
A.~Venturi
P.G.~Verdini
\samepage
\begin{center}
\parbox{15.5cm}{\sl\samepage
Dipartimento di Fisica dell'Universit\`a, INFN Sezione di Pisa,
e Scuola Normale Superiore, I-56010 Pisa, Italy}
\end{center}\end{sloppypar}
\vspace{2mm}
\begin{sloppypar}
\noindent
O.~Awunor,
G.A.~Blair,
G.~Cowan,
A.~Garcia-Bellido,
M.G.~Green,
L.T.~Jones,
T.~Medcalf,
A.~Misiejuk,
J.A.~Strong,
P.~Teixeira-Dias
\nopagebreak
\begin{center}
\parbox{15.5cm}{\sl\samepage
Department of Physics, Royal Holloway \& Bedford New College,
University of London, Egham, Surrey TW20 OEX, United Kingdom$^{10}$}
\end{center}\end{sloppypar}
\vspace{2mm}
\begin{sloppypar}
\noindent
R.W.~Clifft,
T.R.~Edgecock,
P.R.~Norton,
I.R.~Tomalin,
J.J.~Ward
\nopagebreak
\begin{center}
\parbox{15.5cm}{\sl\samepage
Particle Physics Dept., Rutherford Appleton Laboratory,
Chilton, Didcot, Oxon OX11 OQX, United Kingdom$^{10}$}
\end{center}\end{sloppypar}
\vspace{2mm}
\begin{sloppypar}
\noindent
\mbox{B.~Bloch-Devaux},
D.~Boumediene,
P.~Colas,
B.~Fabbro,
E.~Lan\c{c}on,
\mbox{M.-C.~Lemaire},
E.~Locci,
P.~Perez,
J.~Rander,
B.~Tuchming,
B.~Vallage
\nopagebreak
\begin{center}
\parbox{15.5cm}{\sl\samepage
CEA, DAPNIA/Service de Physique des Particules,
CE-Saclay, F-91191 Gif-sur-Yvette Cedex, France$^{17}$}
\end{center}\end{sloppypar}
\vspace{2mm}
\begin{sloppypar}
\noindent
N.~Konstantinidis,
A.M.~Litke,
G.~Taylor
\nopagebreak
\begin{center}
\parbox{15.5cm}{\sl\samepage
Institute for Particle Physics, University of California at
Santa Cruz, Santa Cruz, CA 95064, USA$^{22}$}
\end{center}\end{sloppypar}
\vspace{2mm}
\begin{sloppypar}
\noindent
C.N.~Booth,
S.~Cartwright,
F.~Combley,$^{25}$
P.N.~Hodgson,
M.~Lehto,
L.F.~Thompson
\nopagebreak
\begin{center}
\parbox{15.5cm}{\sl\samepage
Department of Physics, University of Sheffield, Sheffield S3 7RH,
United Kingdom$^{10}$}
\end{center}\end{sloppypar}
\vspace{2mm}
\begin{sloppypar}
\noindent
A.~B\"ohrer,
S.~Brandt,
C.~Grupen,
J.~Hess,
A.~Ngac,
G.~Prange
\nopagebreak
\begin{center}
\parbox{15.5cm}{\sl\samepage
Fachbereich Physik, Universit\"at Siegen, D-57068 Siegen, Germany$^{16}$}
\end{center}\end{sloppypar}
\vspace{2mm}
\begin{sloppypar}
\noindent
C.~Borean,
G.~Giannini
\nopagebreak
\begin{center}
\parbox{15.5cm}{\sl\samepage
Dipartimento di Fisica, Universit\`a di Trieste e INFN Sezione di Trieste,
I-34127 Trieste, Italy}
\end{center}\end{sloppypar}
\vspace{2mm}
\begin{sloppypar}
\noindent
H.~He,
J.~Putz,
J.~Rothberg
\nopagebreak
\begin{center}
\parbox{15.5cm}{\sl\samepage
Experimental Elementary Particle Physics, University of Washington, Seattle,
WA 98195 U.S.A.}
\end{center}\end{sloppypar}
\vspace{2mm}
\begin{sloppypar}
\noindent
S.R.~Armstrong,
K.~Berkelman,
K.~Cranmer,
D.P.S.~Ferguson,
Y.~Gao,$^{13}$
S.~Gonz\'{a}lez,
O.J.~Hayes,
H.~Hu,
S.~Jin,
J.~Kile,
P.A.~McNamara III,
J.~Nielsen,
Y.B.~Pan,
\mbox{J.H.~von~Wimmersperg-Toeller}, 
W.~Wiedenmann,
J.~Wu,
Sau~Lan~Wu,
X.~Wu,
G.~Zobernig
\nopagebreak
\begin{center}
\parbox{15.5cm}{\sl\samepage
Department of Physics, University of Wisconsin, Madison, WI 53706,
USA$^{11}$}
\end{center}\end{sloppypar}
\vspace{2mm}
\begin{sloppypar}
\noindent
G.~Dissertori
\nopagebreak
\begin{center}
\parbox{15.5cm}{\sl\samepage
Institute for Particle Physics, ETH H\"onggerberg, 8093 Z\"urich,
Switzerland.}
\end{center}\end{sloppypar}
}
\footnotetext[1]{Also at CERN, 1211 Geneva 23, Switzerland.}
\footnotetext[2]{Now at Fermilab, PO Box 500, MS 352, Batavia, IL 60510, USA}
\footnotetext[3]{Also at Dipartimento di Fisica di Catania and INFN Sezione di
 Catania, 95129 Catania, Italy.}
\footnotetext[4]{Now at University of Florida, Department of Physics, Gainesville, Florida 32611-8440, USA}
\footnotetext[5]{Also Istituto di Cosmo-Geofisica del C.N.R., Torino,
Italy.}
\footnotetext[6]{Also at Groupe d'Astroparticules de Montpellier, Universit\'{e} de Montpellier II, 34095, Montpellier, France.}
\footnotetext[7]{Supported by CICYT, Spain.}
\footnotetext[8]{Supported by the National Science Foundation of China.}
\footnotetext[9]{Supported by the Danish Natural Science Research Council.}
\footnotetext[10]{Supported by the UK Particle Physics and Astronomy Research
Council.}
\footnotetext[11]{Supported by the US Department of Energy, grant
DE-FG0295-ER40896.}
\footnotetext[12]{Now at Departement de Physique Corpusculaire, Universit\'e de
Gen\`eve, 1211 Gen\`eve 4, Switzerland.}
\footnotetext[13]{Also at Department of Physics, Tsinghua University, Beijing, The People's Republic of China.}
\footnotetext[14]{Supported by the Leverhulme Trust.}
\footnotetext[15]{Permanent address: Universitat de Barcelona, 08208 Barcelona,
Spain.}
\footnotetext[16]{Supported by Bundesministerium f\"ur Bildung
und Forschung, Germany.}
\footnotetext[17]{Supported by the Direction des Sciences de la
Mati\`ere, C.E.A.}
\footnotetext[18]{Supported by the Austrian Ministry for Science and Transport.}
\footnotetext[19]{Now at SAP AG, 69185 Walldorf, Germany}
\footnotetext[20]{Now at Groupe d' Astroparticules de Montpellier, Universit\'e de Montpellier II, 34095 Montpellier, France.}
\footnotetext[21]{Now at BNP Paribas, 60325 Frankfurt am Mainz, Germany}
\footnotetext[22]{Supported by the US Department of Energy,
grant DE-FG03-92ER40689.}
\footnotetext[23]{Now at Institut Inter-universitaire des hautes Energies (IIHE), CP 230, Universit\'{e} Libre de Bruxelles, 1050 Bruxelles, Belgique}
\footnotetext[24]{Also at Dipartimento di Fisica e Tecnologie Relative, Universit\`a di Palermo, Palermo, Italy.}
\footnotetext[25]{Deceased.}
\footnotetext[26]{Now at SLAC, Stanford, CA 94309, U.S.A}
\footnotetext[27]{Now at INFN Sezione di Roma II, Dipartimento di Fisica, Universit\`a di Roma Tor Vergata, 00133 Roma, Italy.}
\footnotetext[28]{Research Fellow of the Belgium FNRS}
\footnotetext[29]{Research Associate of the Belgium FNRS} 
\footnotetext[30]{Now at Liverpool University, Liverpool L69 7ZE, United Kingdom} 
\footnotetext[31]{Supported by the Federal Office for Scientific, Technical and Cultural Affairs through
the Interuniversity Attraction Pole P5/27}   
\setlength{\parskip}{\saveparskip}
\setlength{\textheight}{\savetextheight}
\setlength{\topmargin}{\savetopmargin}
\setlength{\textwidth}{\savetextwidth}
\setlength{\oddsidemargin}{\saveoddsidemargin}
\setlength{\topsep}{\savetopsep}
\normalsize
\newpage
\pagestyle{plain}
\setcounter{page}{1}


\section{Introduction}
The largest part of the cross section for inelastic processes
at LEP2 is due to two-photon scattering.
Interactions of two photons can be studied at $\mathrm{e}^+ \mathrm{e}^-$ 
colliders by investigating the reaction

\begin{equation}
\mathrm{e}^+ \mathrm{e}^- \rightarrow 
\mathrm{e}^+ \mathrm{e}^- \gamma \gamma
\rightarrow \mathrm{e}^+ \mathrm{e}^- \mathrm{X}
  \label{eq:eegg}
\end{equation}
where the photons can be quasi-real or virtual 
and the hadronic final state is denoted by X. 

The analysis reported in this paper focusses on the interactions of virtual 
photons by selecting only double tagged events, i.e., events
where both scattered electrons are detected. Here, electrons and positrons are
generically referred to as electrons. 
Differential cross sections for this process are measured
as a function of several kinematic variables and presented here. 

The purpose of this paper is to compare 
the measured differential cross sections with
the predictions of the PYTHIA \cite{Sjostrand:2000wi} 
and PHOT02 \cite{phot02pap} Monte Carlo
generators, with a next-to-leading-order (NLO) 
QCD calculation \cite{Cacciari:2000cb} and with BFKL 
calculations \cite{Kuraev:1977fs,balitski1978,Kim:1999qq,Kim:1999gp,priv:kim}.

\section{Theoretical Framework}

\begin{figure}[ht!]
  \begin{center}
    \includegraphics[bb=100 400 520 700,width=0.6\textwidth]{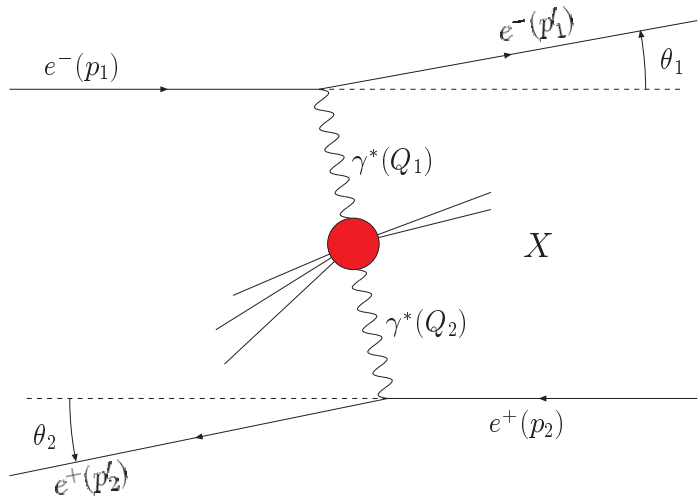}
    \caption{\footnotesize 
      Kinematics of $\gamma\gamma$ interactions at an 
      $\mathrm{e^+e^-}$ collider}
    \label{fig:theo_2}
  \end{center}
\end{figure}

Interactions of virtual photons can be studied 
by requiring that both
electrons be detected after radiating the photons.
The kinematics of these 
electron-induced $\gamma\gamma$ interactions is sketched
in Fig.~\ref{fig:theo_2}. The symbols in
parentheses represent the four-momenta of the particles.

The kinematics can be described by the 
dimensionless Bj\"orken variables of deep inelastic 
scattering:
\begin{eqnarray}
  x_i &=& \frac{Q_i^2}{Q^2_1 + Q^2_2 + W^2_{\gamma \gamma}} \quad,\\
  \label{eq:bjoerken_defx}
  y_i &=& 1 - \frac{E_i^\prime}{E_\mathrm{beam}} 
  \cos^2\left( \frac{\theta_i}{2} \right), 
  \label{eq:bjoerken_defy}
\end{eqnarray}
where $i =1,2$ refers to the scattered electron, positron. 
The hadronic invariant mass $W_{\gamma \gamma}$ used in these
definitions is
obtained from the energies $E_h$ and the momenta $\vec{p}_h$ of
the final state particles $h$ by 
\begin{equation}
  W_{\gamma \gamma}^2 = 
  \left( \sum_h E_h \right)^2 - 
  \left( \sum_h \vec{p}_h \right)^2 =
  E^2_\mathrm{X} - \vec{p}^{\;2}_\mathrm{X} \ .
\label{eq:wggcalc}
\end{equation}
The virtualities of the photons $Q^2_i$
are 
\begin{equation}
  Q_i^2 = -(p_i -p_i^\prime)^2 = 2 E_\mathrm{beam} E_i^\prime
(1-\cos\theta_i)
  \label{eq:q2def}
\end{equation}
for $\theta_i \gg m_\mathrm{electron} / E_i$, where 
$\theta_i$ are the scattering angles of the deflected leptons. 

From Eq.~(\ref{eq:q2def}) it is possible to
select interactions of virtual photons by requiring
that the scattered electrons be detected at large angles.
The accessible range of virtualities and therefore the phase space
depends on the region where electrons can be detected. 

For the comparison of the data with BFKL 
calculations the following quantity is defined, 
\begin{equation}
  Y = \log{\left( \frac{s}{s_0} \right)} \ , 
  \label{eq:Y_def}
\end{equation}
where $\sqrt{s}$ is the LEP centre-of-mass energy and 
$s/s_0 = s y_1 y_2 / \sqrt{Q^2_1 Q^2_2} 
\approx 
W^2_{\gamma \gamma} / \sqrt{Q^2_1 Q^2_2}$, the approximation requiring 
$W^2_{\gamma \gamma} \gg Q^2_i$. 

To correct the data for detector effects, 
the signal Monte Carlo simulation for this analysis uses 
PYTHIA 6.151 \cite{Sjostrand:2000wi,Friberg:2000ra,Friberg:2000nx}
and as an alternative PHOT02 \cite{phot02pap}. 
The PHOJET generator \cite{phojet}, which simulates 
untagged events, gives information for further systematic studies.

The PHOT02 generator is a
combination of different two-photon generators.
The main contribution to this analysis comes from the QED part, based 
on a program  
that gives a matrix element calculation for 
$\mathrm{e}^+ \mathrm{e}^- \rightarrow \mathrm{e}^+ \mathrm{e}^- 
\ell^+ \ell^- $, where $\ell^\pm$ 
are charged fermions 
\cite{Vermaseren:1983cz,Cochard:1980iy,Bhattacharya:1977re}. 
A small contribution arises from
the vector-meson dominance model (VDM)
\cite{Ginzburg:1982bs,Bonneau:1973kg}.
The implementation of the $\gamma\gamma$ generator in PYTHIA is based 
on a model described in 
\cite{Friberg:2000ra,Friberg:2000nx}.

The background from leptonic double tagged two-photon events
is also simulated with PHOT02. For annihilation events three generators are 
employed: KORALZ \cite{koralz}, PYTHIA and HERWIG \cite{Marchesini:1992ch}.
The latter is used to generate single
tagged two-photon events contributing to the background.

\section{ALEPH Detector}

A detailed description of the ALEPH detector and its performance
can be found in
Ref.~\cite{ALEPHdetperfvdet}. 
The inner part of the ALEPH detector is
dedicated to the reconstruction of the trajectories of charged
particles with a two-layer silicon strip
vertex detector (VDET), a cylindrical drift chamber (ITC) and a large time
projection chamber (TPC). The three tracking detectors are immersed in a
$1.5\unit{T}$ axial magnetic field provided by a superconducting solenoidal
coil. Together they measure charged particle 
momenta with a resolution
of $\delta p_{\mathrm t}/p_{\mathrm t} = 6 \times 10^{-4} p_{\mathrm t}
\oplus 0.005$ ($p_{\mathrm t}$ in GeV/$c$). 

Photons are identified in the electromagnetic calorimeter
(ECAL), situated between the TPC and the coil. The ECAL
is a lead/proportional-tube
sampling calorimeter segmented into $0.9^{\circ} \times 0.9^{\circ}$ projective
towers read out in three sections in depth. It has a total thickness
of 22 radiation lengths and yields a relative energy resolution of
$0.18/\sqrt{E} + 0.009$ ($E$ in GeV) for isolated photons. Electrons 
crossing the TPC are
identified by their transverse and longitudinal shower profiles in
ECAL and their specific ionization in the TPC.

The iron return yoke is instrumented with 23 layers of streamer tubes and
forms the hadron calorimeter (HCAL). The latter provides a relative
energy resolution for charged and neutral hadrons of $0.85/\sqrt{E}$ 
($E$ in GeV). 
Muons are distinguished from hadrons by their characteristic
pattern in HCAL and by the muon chambers, which are composed of two 
double-layers of streamer tubes outside HCAL.

The information from the tracking detectors and the calorimeters are
combined in an energy-flow algorithm~\cite{ALEPHdetperfvdet}. 
For each event, the
algorithm provides a set of charged and neutral reconstructed particles,
called energy-flow objects.

Two small-angle luminosity calorimeters, the silicon luminosity 
calorimeter (SICAL) and the luminosity calorimeter (LCAL), are used to detect 
and measure the energies of the electrons from beam-beam scattering 
including the electrons in the final state of reaction~(1). 
The SICAL uses 12 silicon/tungsten layers to sample showers. 
It is mounted around the beam pipe and covers angles 
from $34$ to $58\unit{mrad}$. An energy resolution of 
$0.225\sqrt{E}$ ($E$ in GeV) is achieved.
The LCAL 
is a lead/proportional-tube calorimeter, similar to ECAL, 
placed around the beam pipe 
at each end of the ALEPH detector. It monitors angles from $45$ 
to $160\unit{mrad}$ with an energy resolution of $0.33\sqrt{E}$ ($E$ in GeV). 

\section{Data Sample}

\subsection{Event Selection}

The analysis is based on the data taken with the ALEPH detector
from 1998 to 2000 at centre-of-mass energies 
$\sqrt{s} = 188 - 209\,\mathrm{GeV}$ and corresponds to 
an integrated luminosity 
of $640\,\mathrm{pb}^{-1}$.

The event selection is performed in three stages:
detection of scattered electrons, verification of the presence of a
hadronic system, and background reduction.

\begin{itemize}
\item Detection of scattered electrons.\\
The luminosity detectors SICAL and LCAL detect the 
scattered electrons. Thus the polar angular range is restricted
to $35\,\mathrm{mrad} < \theta_i <  55\,\mathrm{mrad}$ (SICAL)
and $60\,\mathrm{mrad} < \theta_i < 155\,\mathrm{mrad}$ (LCAL). 
The energy threshold is set to $E_i' > 0.3\,E_\mathrm{beam}$.
\item Verification of the hadronic system.\\
To ensure that the final state is a hadronic system and not
a lepton pair, at least three charged particles are required.
The visible mass $W_{\gamma \gamma}$ of the hadronic system 
must be larger than 3\,GeV$/c^2$.
\item Background reduction.\\
The total visible energy 
$E_\mathrm{tot} = E_1^\prime + E_2^\prime + E_{\mathrm{X}}$
must be larger than 70\% of the nominal centre-of-mass energy. 
To reject remaining Bhabha events the acolinearity 
of the scattered leptons is required to be less than $179.5^{\circ}$.
\end{itemize}

\subsection{Backgrounds}

With these cuts 891 events were selected in the data with
206.1 expected background events.
The three remaining sources of background are the following.
\begin{itemize}
\item Double-tagged leptonic $\gamma\gamma$ events containing mostly 
$\tau^+$$\tau^-$ as 
estimated with the PHOT02 Monte Carlo simulation.
\item Superpositions 
of single tagged $\gamma \gamma$ events and
off-momentum electrons.
In order to appraise this background source it is necessary to
extract the probability of finding an off-momentum electron
in an arbitrary data event. This is done by
looking for additional energy deposits from off-momentum 
electrons in Bhabha events. These off-momentum electrons are then added 
to single tagged events simulated with the HERWIG Monte Carlo or alternatively
to those taken from data.
\item Annihilation events ($\mathrm{e}^+ \mathrm{e}^- \rightarrow 
\mathrm{q \bar{q}}$) which 
can also fake the topology of double tagged events.
This background source, dominated by radiative Z return, 
is generated using the KORALZ Monte Carlo. 
\end{itemize}

The trigger efficiency is estimated by comparing the
rates of two independent triggers: the Bhabha event trigger
and the combination of non-Bhabha (charged-track and neutral-energy)
triggers. The selected events are found to allways fulfill the two 
trigger conditions. Therefore, the trigger is taken to be $100\%$ 
with an uncertainty of $2\%$ and no correction is applied. 

\subsection{Selection Results}

The numbers of events obtained in data, the signal Monte Carlo
simulations 
and the estimated backgrounds are
summarized in Table~\ref{tab:cutres}.
The total cross section of the signal Monte Carlo simulations
was normalized to data after background subtraction. 
The cross section of the PYTHIA Monte Carlo generator was reduced by 
12\% and the cross section extracted from PHOT02 was increased by 30\%. 

\begin{table}[t]
  \begin{center}
    \caption{\footnotesize
      Numbers of observed events for combinations of detectors.
      The notation ``MC + back'' stands for PYTHIA Monte Carlo plus all
      background sources. PHOT02 is given for comparison as an alternative 
      signal Monte Carlo.
      }
\vspace*{3mm}
    \begin{tabular}{c c c c c c c c}
      \rotatebox{60}{detector} & \rotatebox{60}{data}
        & \rotatebox{60}{MC + back} & \rotatebox{60}{PYTHIA} 
        & \rotatebox{60}{PHOT02}
        & \rotatebox{60}{$\gamma \gamma \rightarrow \tau\tau$} 
        & \rotatebox{60}{$\mathrm{e^+ e^- \rightarrow q \bar{q}}$} &
      \rotatebox{60}{off momentum} \\
 \hline
      SiCAL - SiCAL & 243 & 195 & 150 & 177 &  13 &  \phantom{0}3 &  30 \\
      SiCAL - LCAL  & 388 & 447 & 360 & 328 &  37 &  30 &  21 \\
      LCAL - LCAL   & 260 & 249 & 176 & 179 &  23 &  51 &  \phantom{0}0 \\
      total         & 891 & 891 & 685 & 685 &  73 &  83 &  50 \\
 \hline

    \end{tabular}
    \label{tab:cutres}
  \end{center}
\end{table}

\begin{figure}[!htp]
  \begin{center}
    \includegraphics[width=0.45\textwidth]{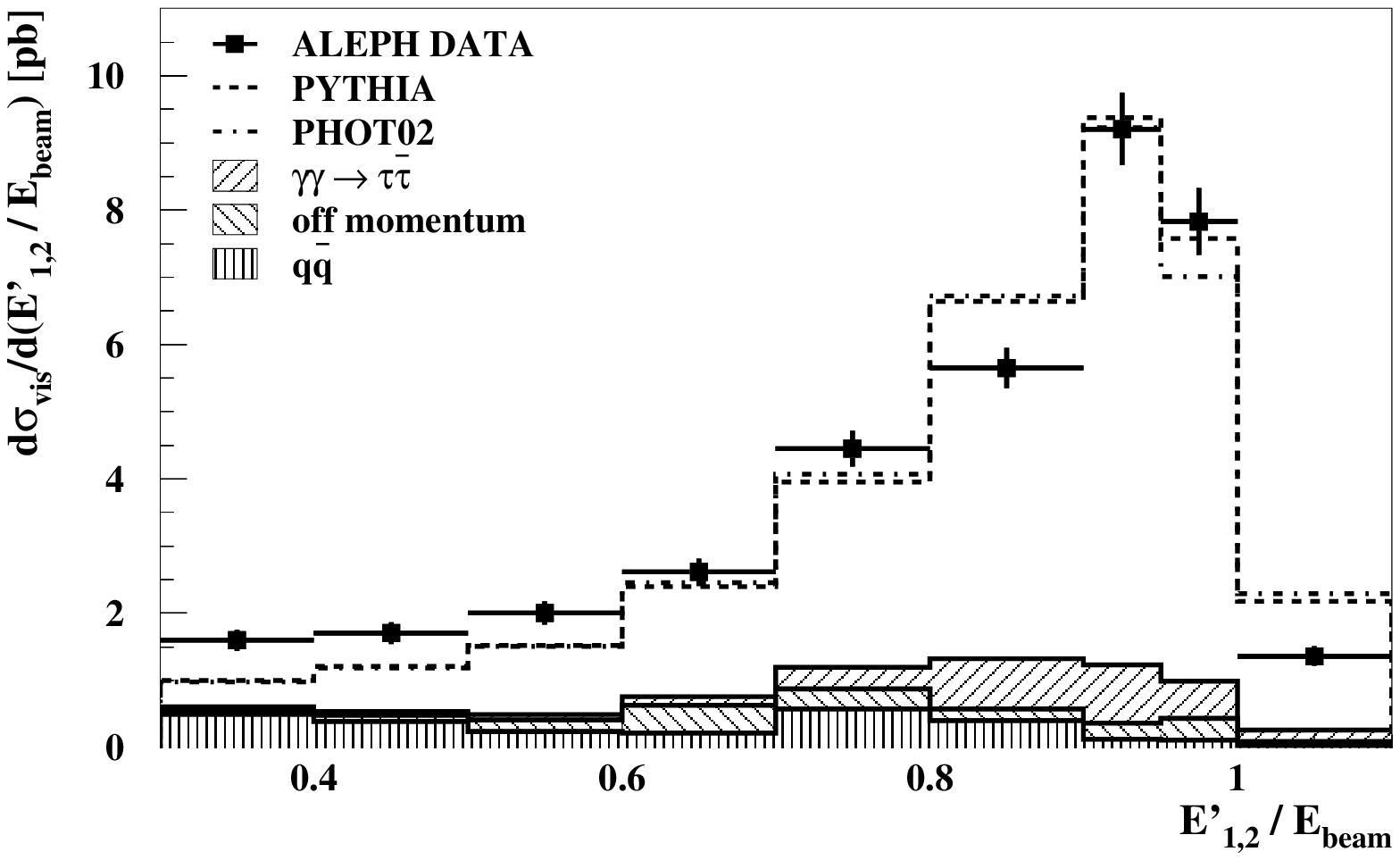}
    \includegraphics[width=0.45\textwidth]{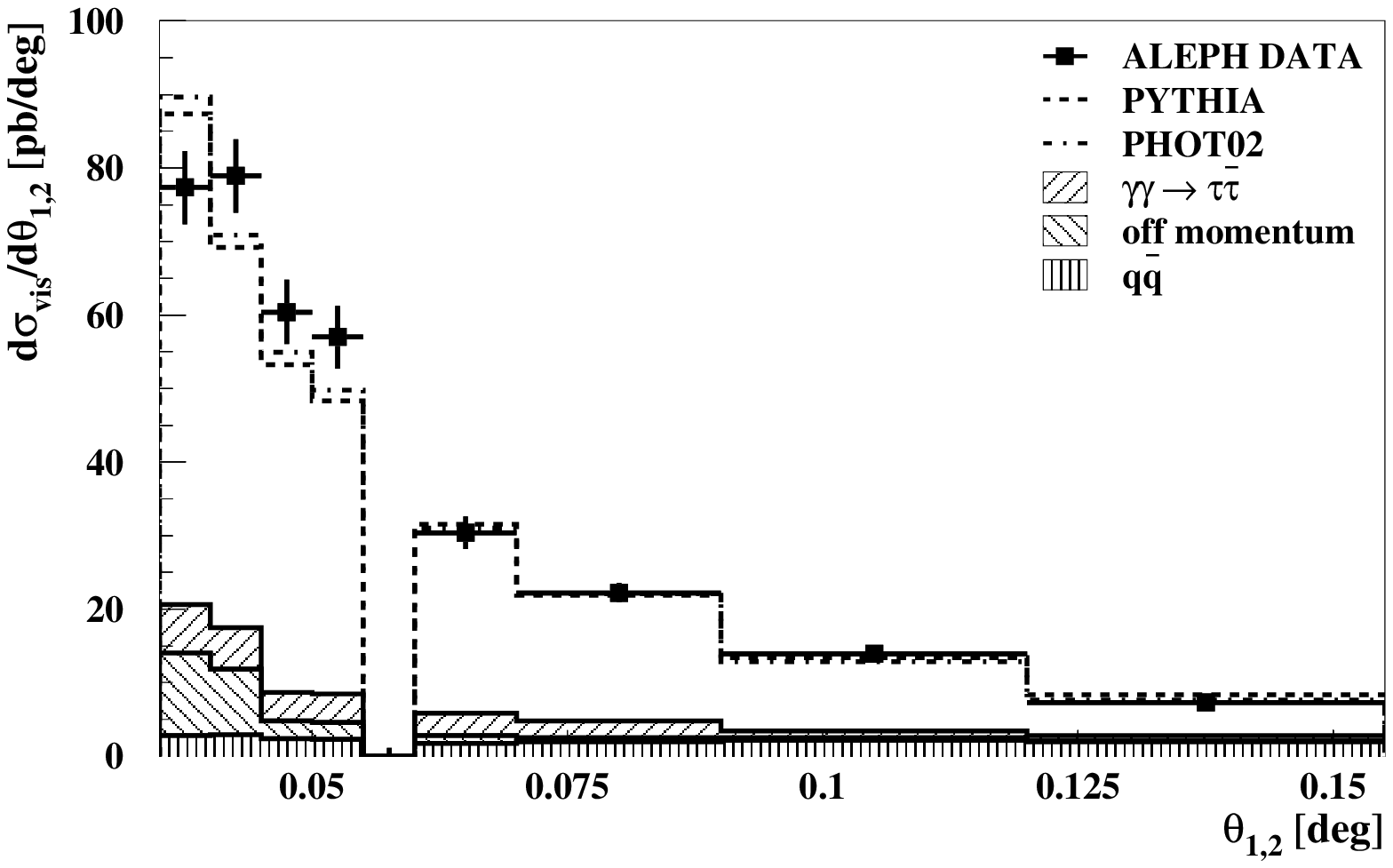}
    \includegraphics[width=0.45\textwidth]{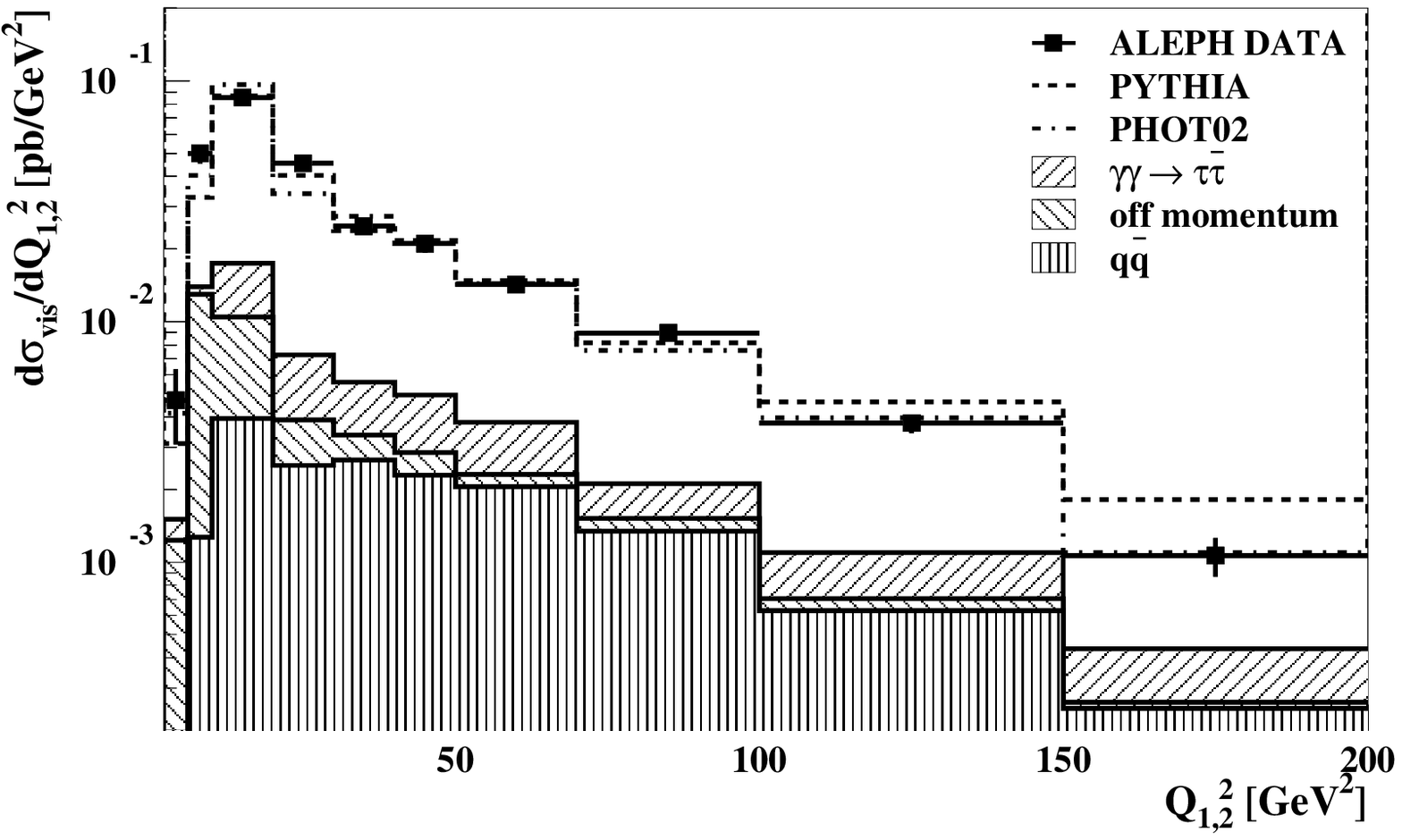}
    \includegraphics[width=0.45\textwidth]{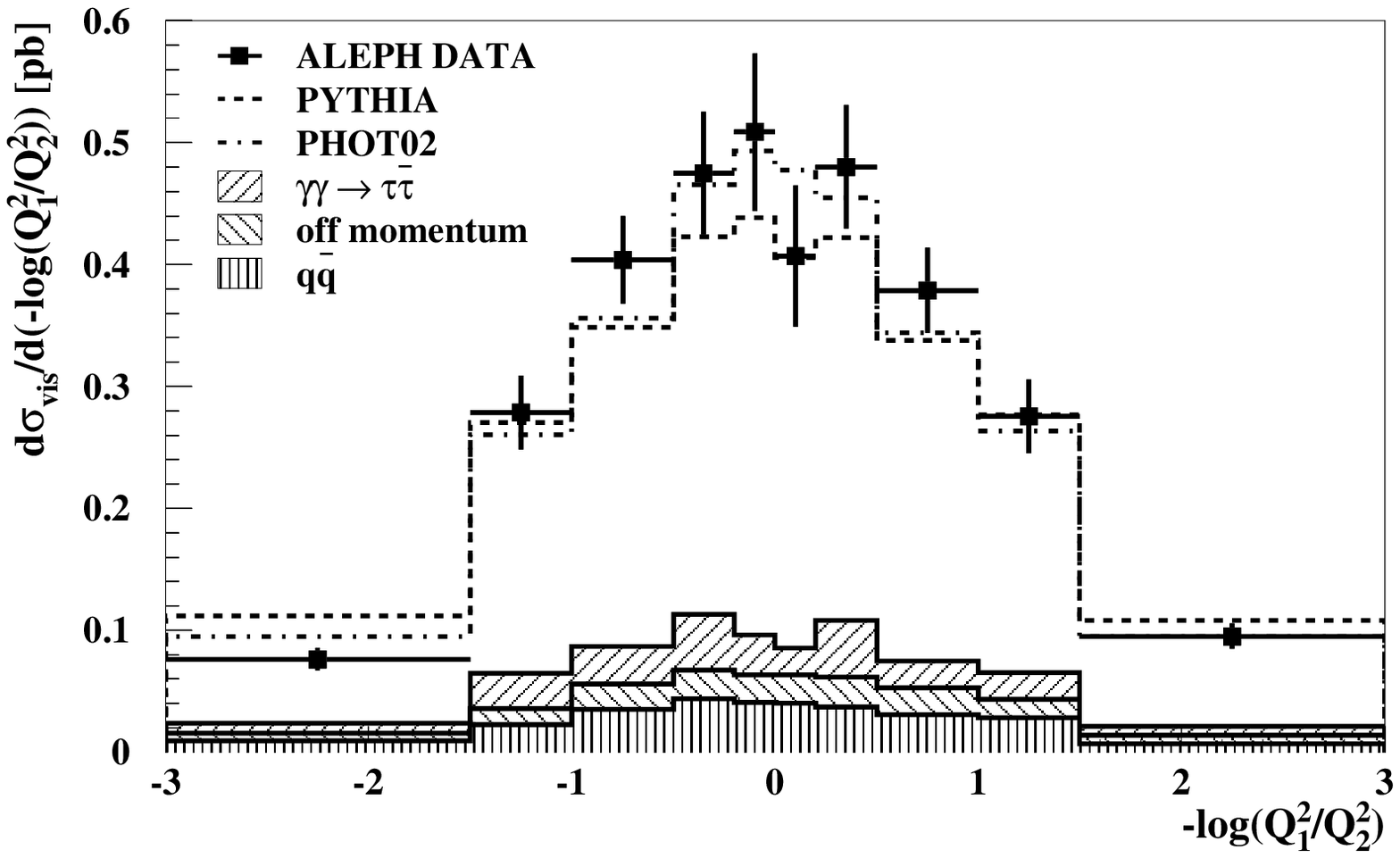}
    \includegraphics[width=0.45\textwidth]{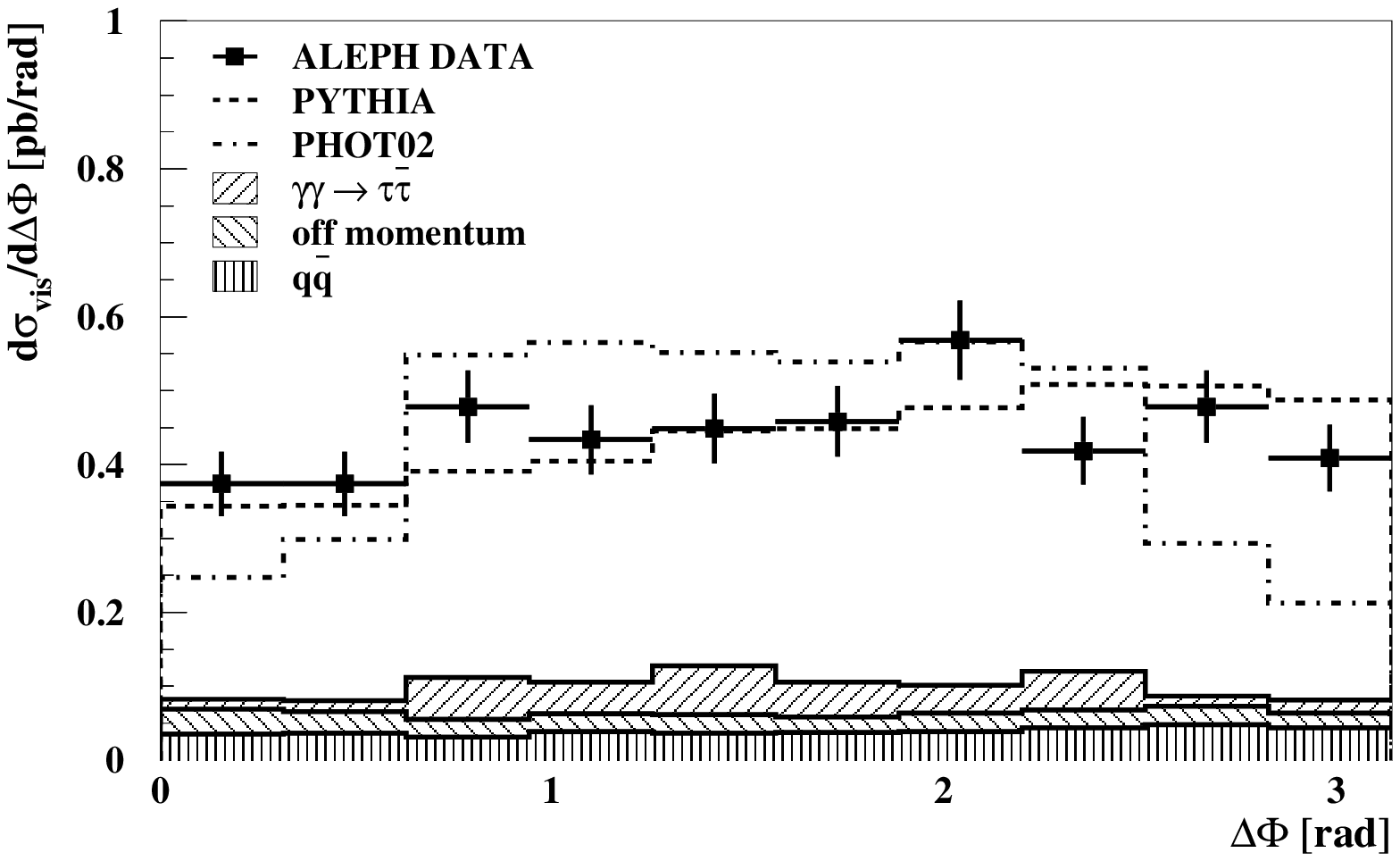}
    \includegraphics[width=0.45\textwidth]{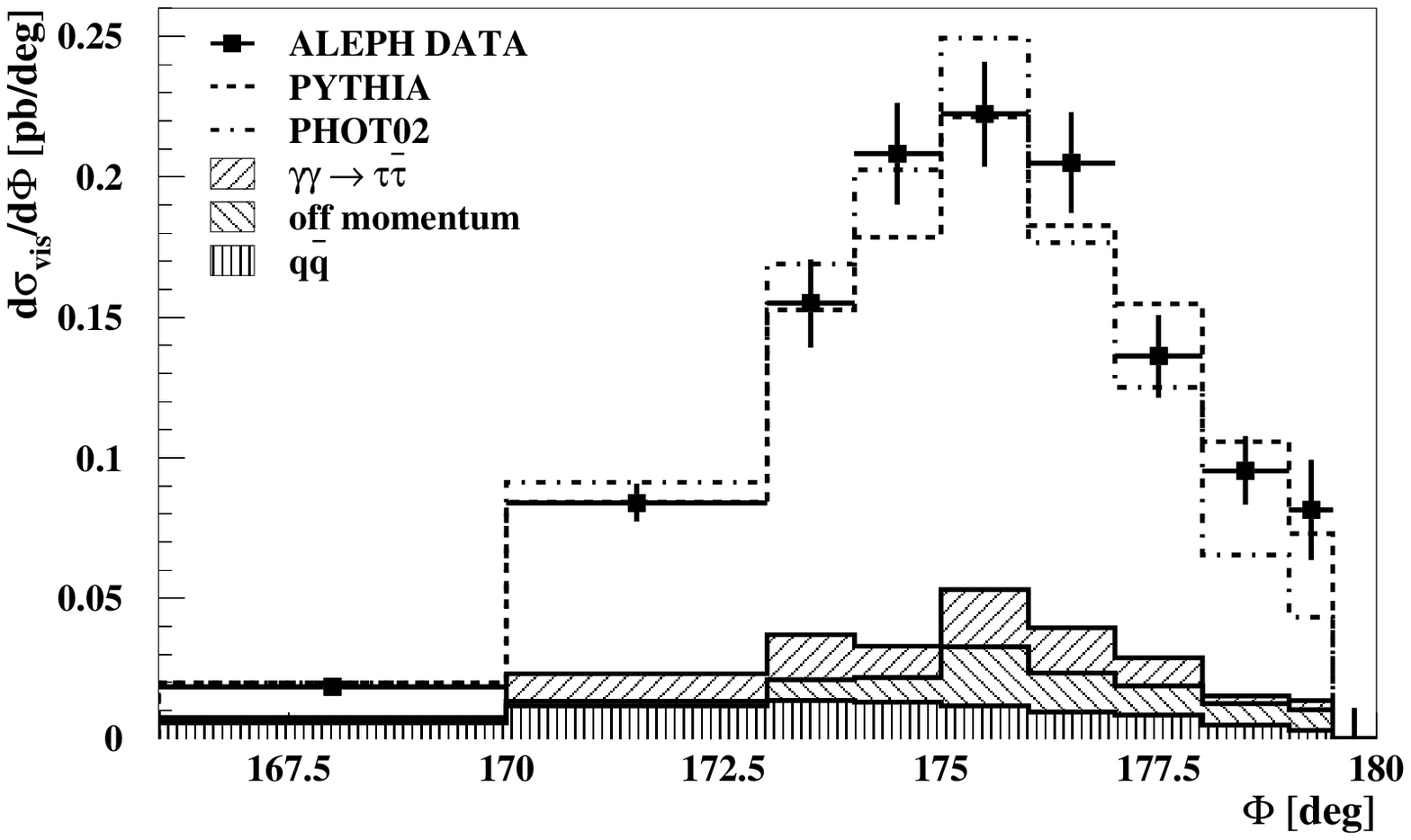}
    \includegraphics[width=0.45\textwidth]{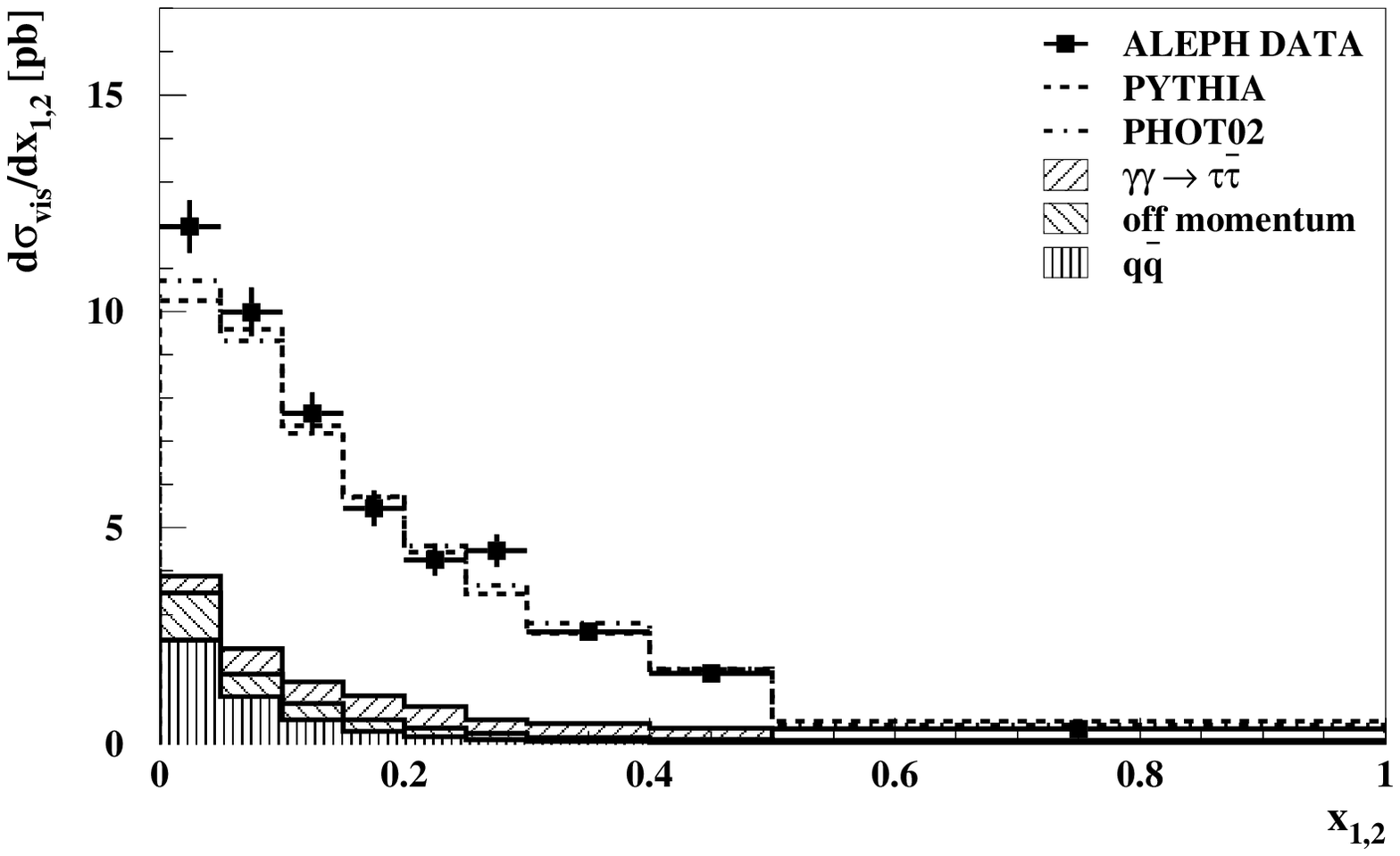}
    \includegraphics[width=0.45\textwidth]{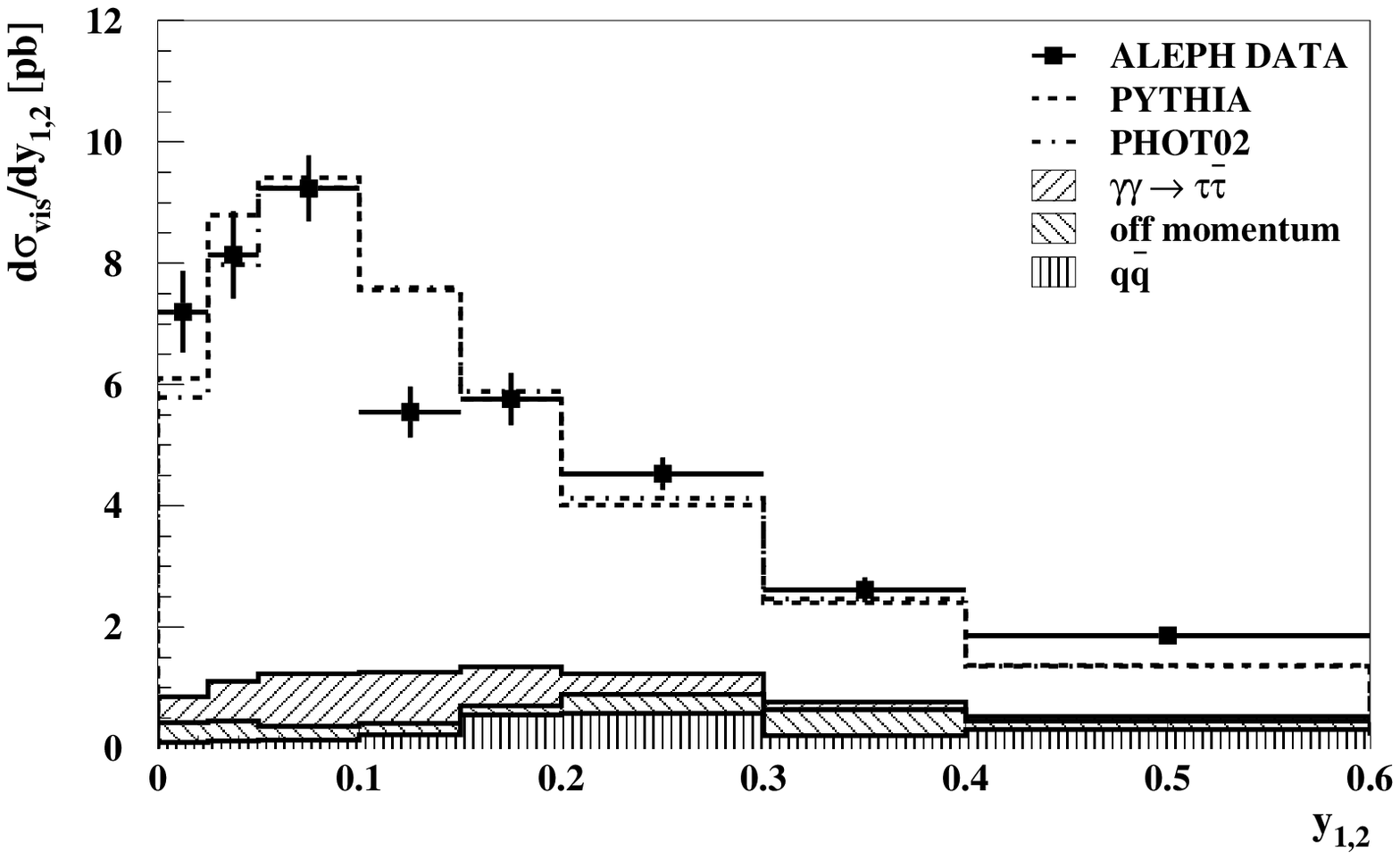}
    \caption{\footnotesize
    Differential cross sections as a function of various observables.
      Shown are (from top left to bottom right):
The relative tag energy $E^\prime_\mathrm{1,2} / E_\mathrm{beam}$;
%
the polar angle of the scattered electrons $\theta_\mathrm{1,2}$;
%
the virtualities $Q^2_i$ of the photons;
%
the ratio of the two virtualities, shown as $\Delta Q = -\log Q^2_1 /
  Q^2_2$;
%
the acoplanarity angle $\Delta\phi$ between the scattered electrons; 
%
the acolinearity $\Phi$ between the scattered electrons;
%
the deep inelastic scattering variables $x_i$ and $y_i$.
The plots contain the background contributions (filled areas), the 
PYTHIA and PHOT02 Monte Carlo predictions plus background (lines), and 
the data (points) with the statistical errors.
}
    \label{fig:vis1}
  \end{center}
\end{figure}

Several measured spectra are given in Figs.~\ref{fig:vis1} and
\ref{fig:vis2} showing the comparison between data and simulations 
after normalization.

\section{Acceptance Corrections and Systematic Errors}

A simple bin-by-bin
method was applied to correct for detector inefficiencies.
The correction factors were calculated for each bin as 
\begin{equation}
  \varepsilon_\mathrm{bin} = \frac{N_{\mathrm{true}}}{N_{\mathrm{detected}}}
  \label{eq:korr}
\end{equation}
where $N_{\mathrm{true}}$ is the number of generated events in a given
bin and $N_{\mathrm{detected}}$ the number of detected events in the
same bin according to the simulation.
The corrected data values for a bin, $R_{\mathrm{cor}}$, were then
derived from the measured values, $R_\mathrm{visible}$, by

\begin{equation}
  R_{\mathrm{cor}} = 
  (R_\mathrm{visible} - R_\mathrm{background}) \varepsilon_\mathrm{bin}
  \label{eq:korrdat}
\end{equation}
where $R_\mathrm{background}$ is the expected background.

\begin{figure}[!htp]
  \begin{center}
    \includegraphics[width=0.45\textwidth]{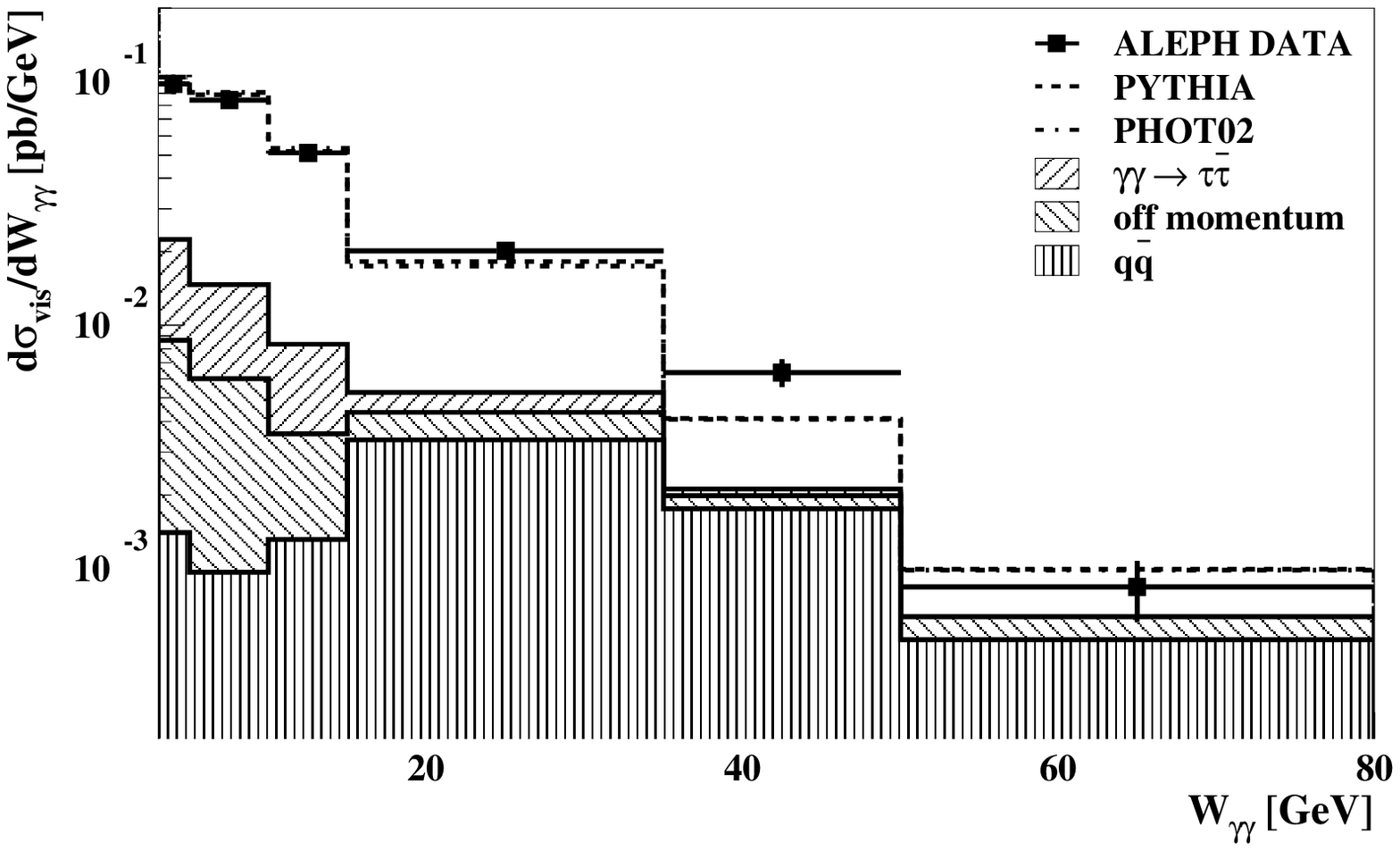}
    \includegraphics[width=0.45\textwidth]{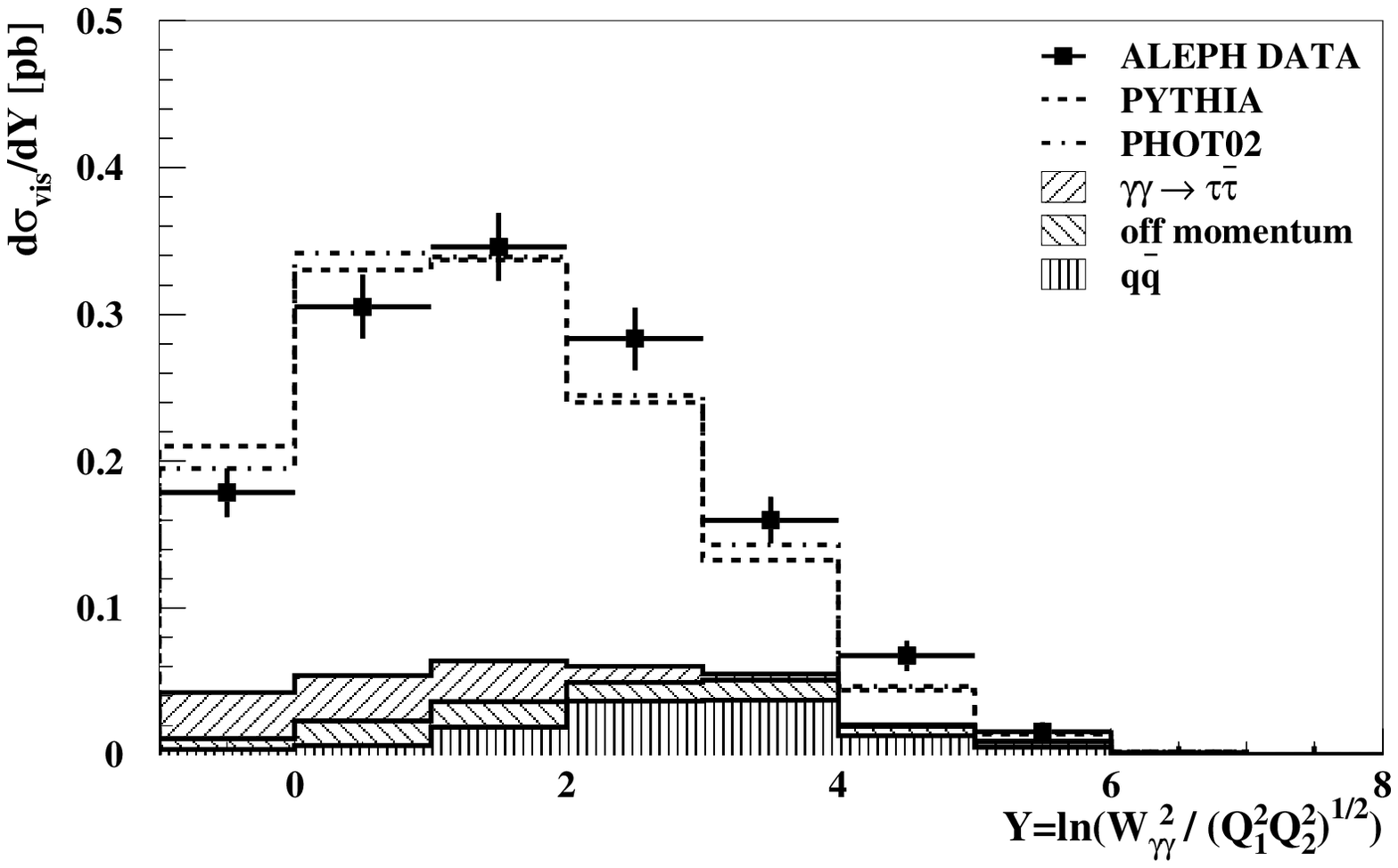}
    \includegraphics[width=0.45\textwidth]{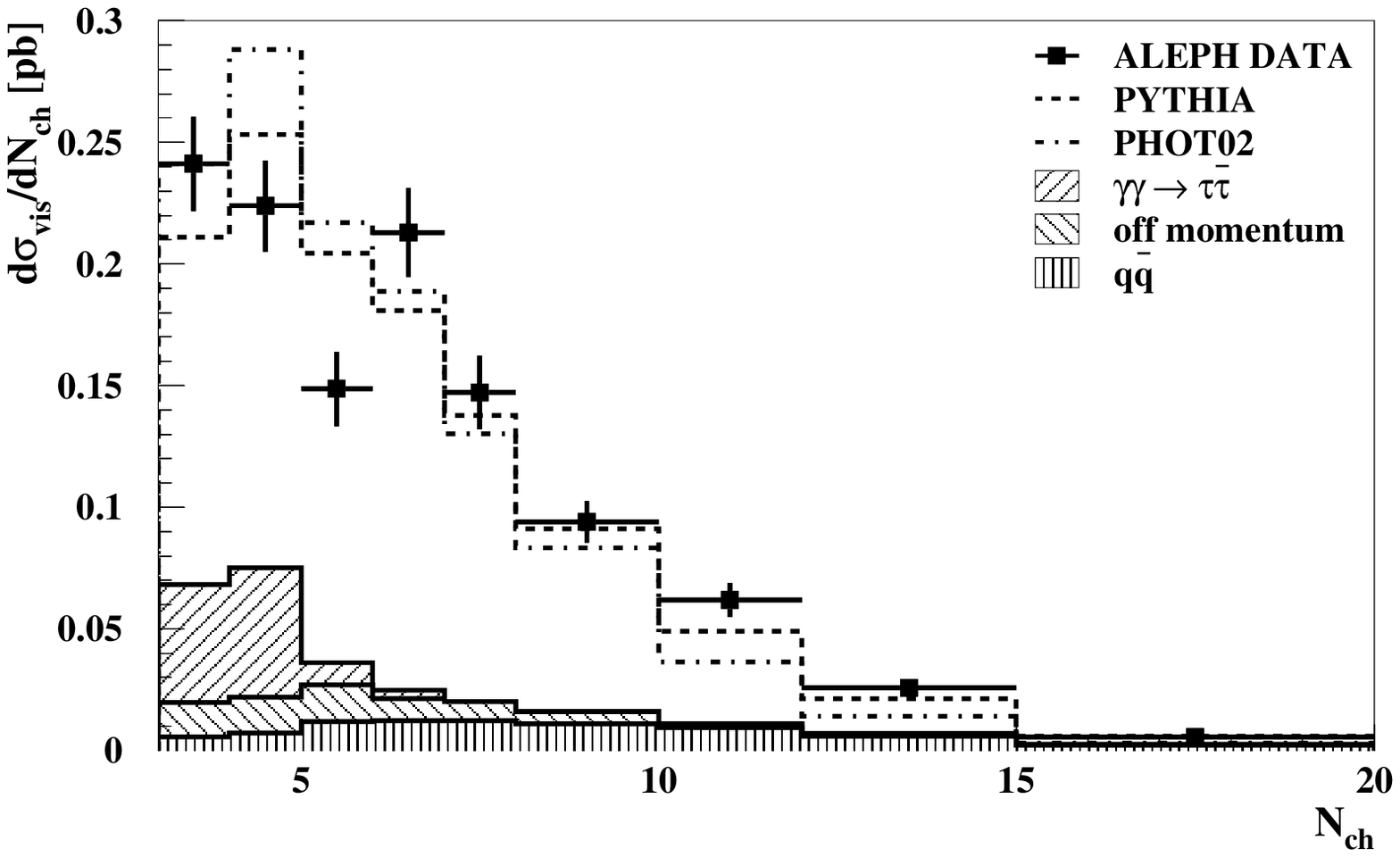}
    \includegraphics[width=0.45\textwidth]{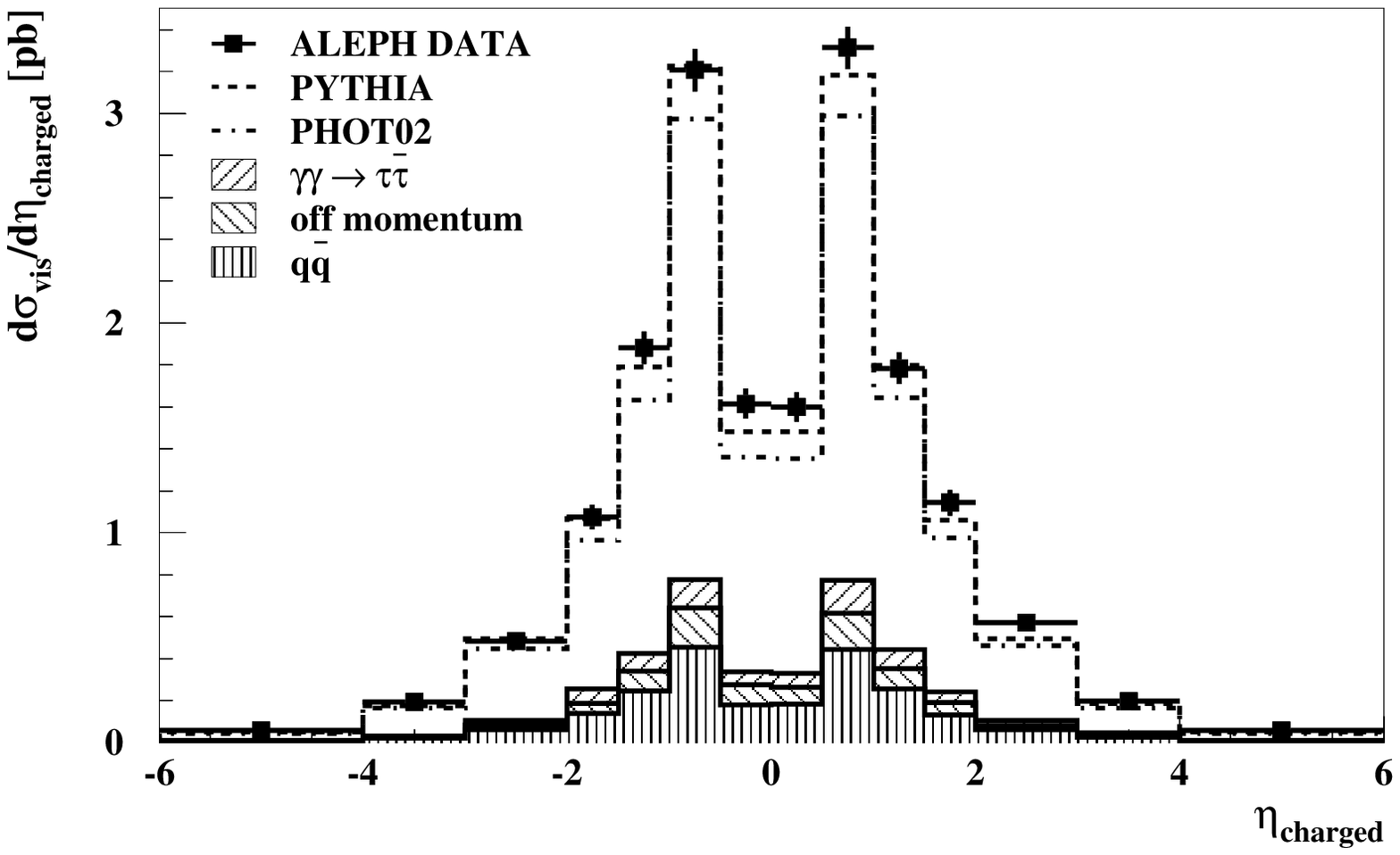}
    \includegraphics[width=0.45\textwidth]{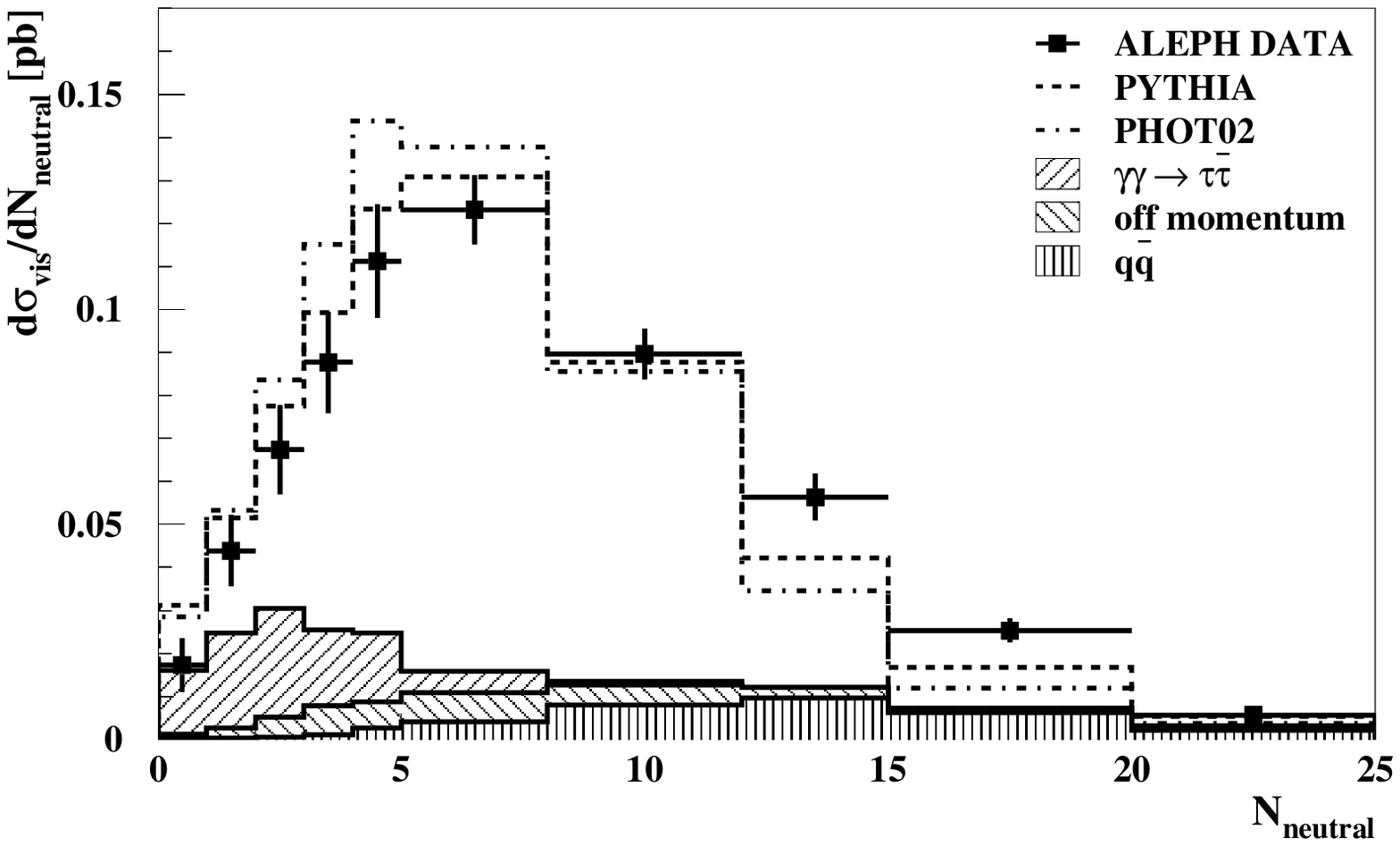}
    \includegraphics[width=0.45\textwidth]{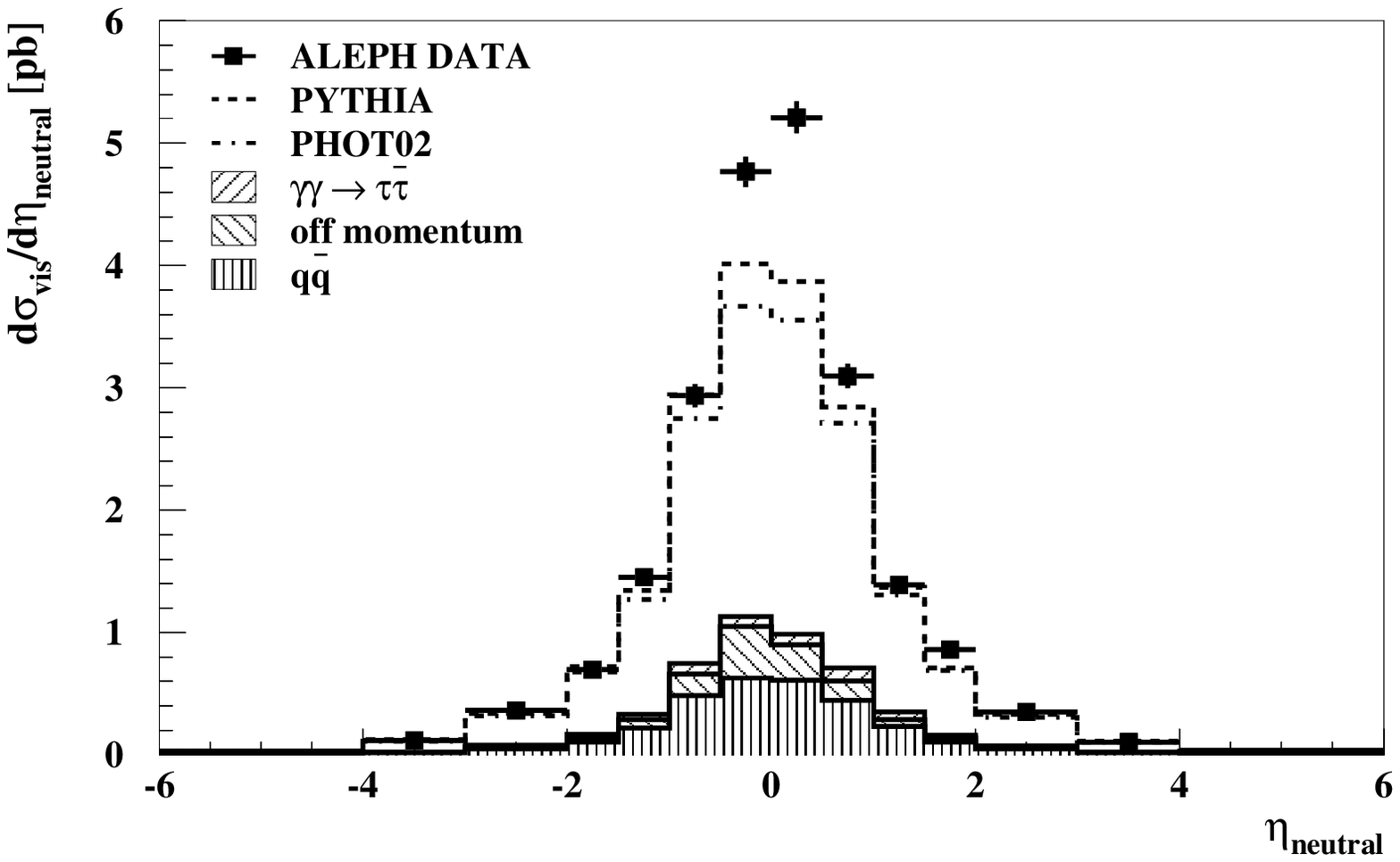}
    \includegraphics[width=0.45\textwidth]{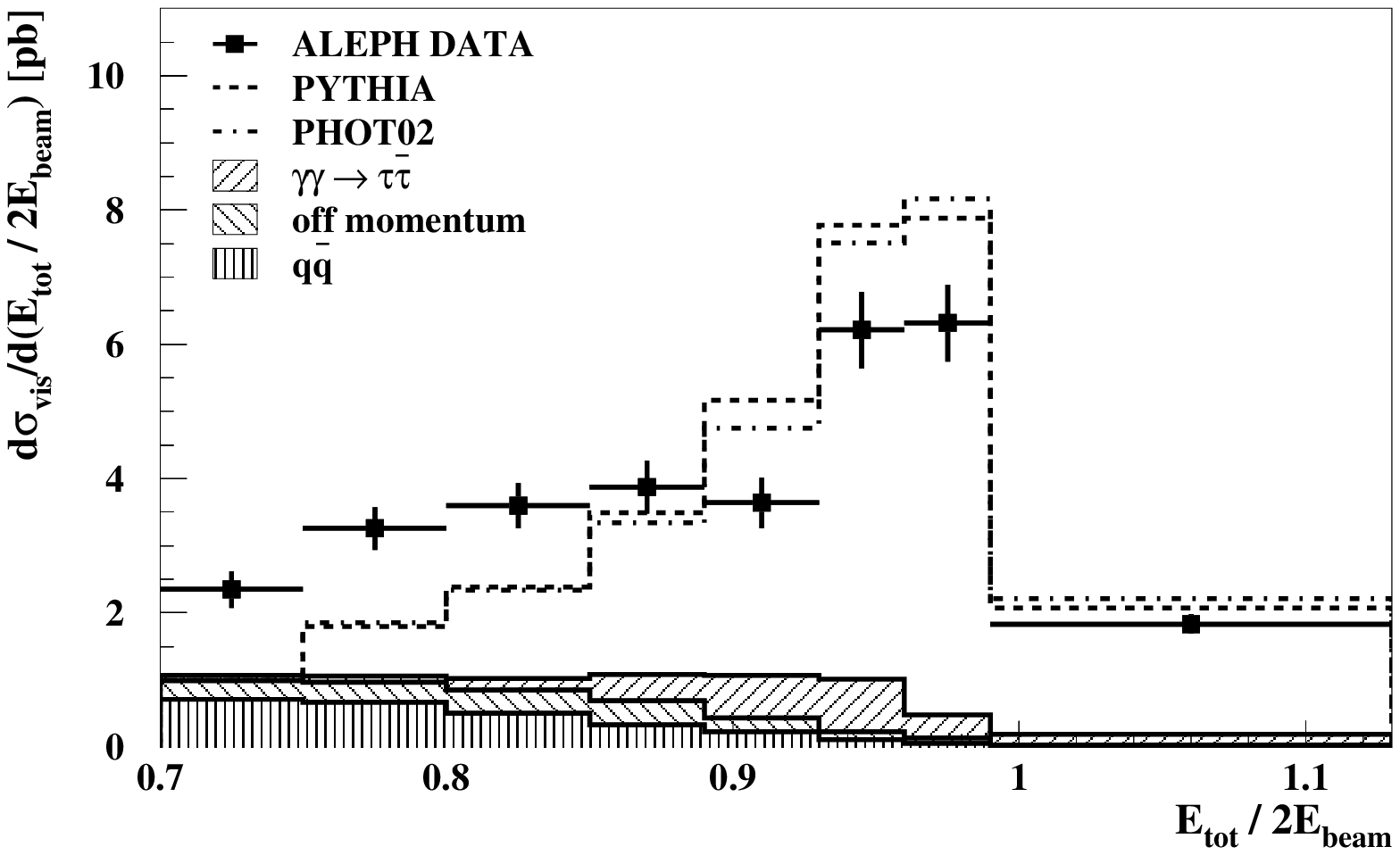}
    \includegraphics[width=0.45\textwidth]{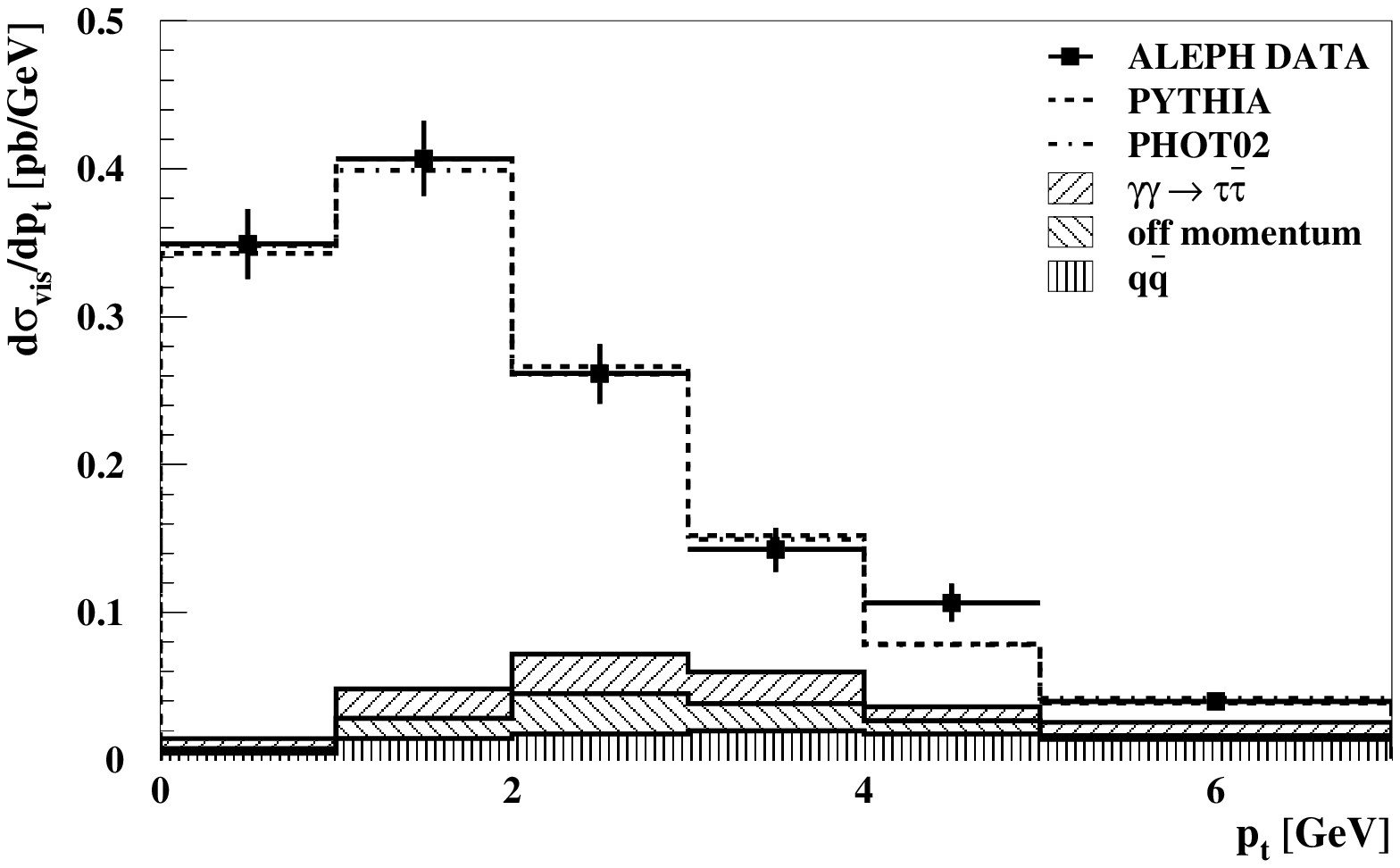}
    \caption{\footnotesize
    Differential cross sections as a function of various observables.
      Shown are (from top left to bottom right): 
%
The mass of the hadronic system $W_{\gamma\gamma}$;
%
the quantity $Y = \log W_{\gamma\gamma}^2 / \sqrt{Q^2_1 Q^2_2}$;
%
the number of charged tracks $N_\mathrm{ch}$;
%
the pseudorapidity 
  $\eta_\mathrm{ch} = -\log \tan (\theta_\mathrm{ch} / 2)$ of 
  the charged tracks;
%
the number of neutral objects $N_\mathrm{neutral}$;
%
the pseudorapidity $\eta_\mathrm{neutral}$ of the neutral
objects;
%
the relative total visible energy $E_\mathrm{tot} / 2 E_\mathrm{beam}$;
%
the total transverse momentum $p_\mathrm{t}$ of the event with respect to
  the beam pipe.
The plots contain the background contributions (filled areas), the 
PYTHIA and PHOT02 Monte Carlo predictions plus background (lines), and 
the data (points) with the statistical errors.
}
    \label{fig:vis2}
  \end{center}
\end{figure}

In order to estimate the systematic effect caused by an imperfect
detector simulation all energy resolutions were varied by 10\% 
(Fig.~\ref{fig:vis2}). 
 
The uncertainty due to a possible shift in the energy scale of the
luminosity monitors SICAL and LCAL (Fig.~\ref{fig:etag}) was estimated 
by introducing an offset of 0.5\,GeV to the measured energy.
The polar and azimuthal angles of the electrons
were shifted by 0.25\,mrad and 0.5\,mrad. These seemingly large
adjustments were made to estimate a possible systematic uncertainty
due to a poor description
of small polar angles of the electrons (Fig.~\ref{fig:theta_tag}).

The cross sections of the background processes were changed 
conservatively by $\pm10\%$.

The PHOT02 Monte Carlo simulation was used instead of PYTHIA to
correct for detector effects. In both cases the systematic uncertainties
due to
statistical fluctuations in the Monte Carlo sample were small.

Systematic differences between the data collected in
different years were
not observed. 

The systematic error was computed as the quadratic sum of the various 
contributions. 
The main contributions to the systematic error come from varying the
energy resolutions and the tag energy.

\section{Results}

The measured cross sections, corrected for detector effects, are given
in Figs.~\ref{fig:etag} to \ref{fig:Y}. The corresponding bin-contents 
are given in the Appendix. 
The results are compared with the PYTHIA and PHOT02 Monte Carlo models and 
the NLO QCD prediction~\cite{Cacciari:2000cb}. 

Figure~\ref{fig:etag} 
shows the cross section
as a function of the tag energy. The cross section predicted by the
NLO QCD calculation comes out too low by $20\%$. The tail towards low
energies is slightly underestimated by all models.
The polar angle $\theta$ of the scattered electrons is given in
Fig.~\ref{fig:theta_tag}.
The first
measured point (35 to 40\,mrad) is overestimated by the Monte Carlo. 
The gap between SICAL and LCAL (55 to 60\,mrad) 
is interpolated for the further results in Figs.~\ref{fig:q2_tag} 
to \ref{fig:Y}. 

The virtuality of the photons $Q^2_i$ are well simulated over the
whole accessible range (Fig.~\ref{fig:q2_tag}).
This also applies to the ratio of the
two virtualities (Fig.~\ref{fig:q1q2_tag}).

The acoplanarity angle $\Delta \phi$ (Fig.~\ref{fig:DPHI_tag}) 
and the acolinearity $\Phi$ between the scattered electrons 
(Fig.~\ref{fig:Phi_tag}) are 
well described by the PYTHIA
prediction. The NLO QCD calculation yields a slightly too 
low cross section at
large angles, and PHOT02 does not describe well the acoplanarity angle 
distribution. 
The mass of the hadronic system is again well reproduced by the Monte
Carlo simulations. The NLO QCD calculation comes out with a slightly
too low cross section for small masses 
(Fig.~\ref{fig:Wgg}).
The deep inelastic scattering variables $x_i$ 
(Fig.~\ref{fig:x})
and $y_i$ 
(Fig.~\ref{fig:y})
in data are well described by 
the Monte Carlo simulations.
The NLO QCD calculation fails in reproducing the cross section as a
function of $x_i$, while $y_i$ agress well.
The cross section as a function of 
$Y$ is given in Fig.~\ref{fig:Y}. 
Again the Monte Carlo simulations describe the measured spectrum well.

Finally the data are compared to BFKL predictions.
The calculation was done for photons with virtualities
of $12\,\mathrm{GeV}^2$ (90\% of the data) 
and $38\,\mathrm{GeV}^2$ (10\% of the data).
Since this calculation assumes that both photons have the same
virtuality, a further cut was added to the event selection \cite{priv:carlo}:
\begin{equation}
  | \Delta Q | = | \log Q^2_1 /  Q^2_2 | < 1.0 \ \ .
  \label{eq:bfklcut}
\end{equation}
The whole analysis, including systematic error estimation, was redone 
with this new cut, which reduced the statistics by 40\%.

\begin{figure}[!htp]
  \begin{center}
    \includegraphics[width=0.85\textwidth]{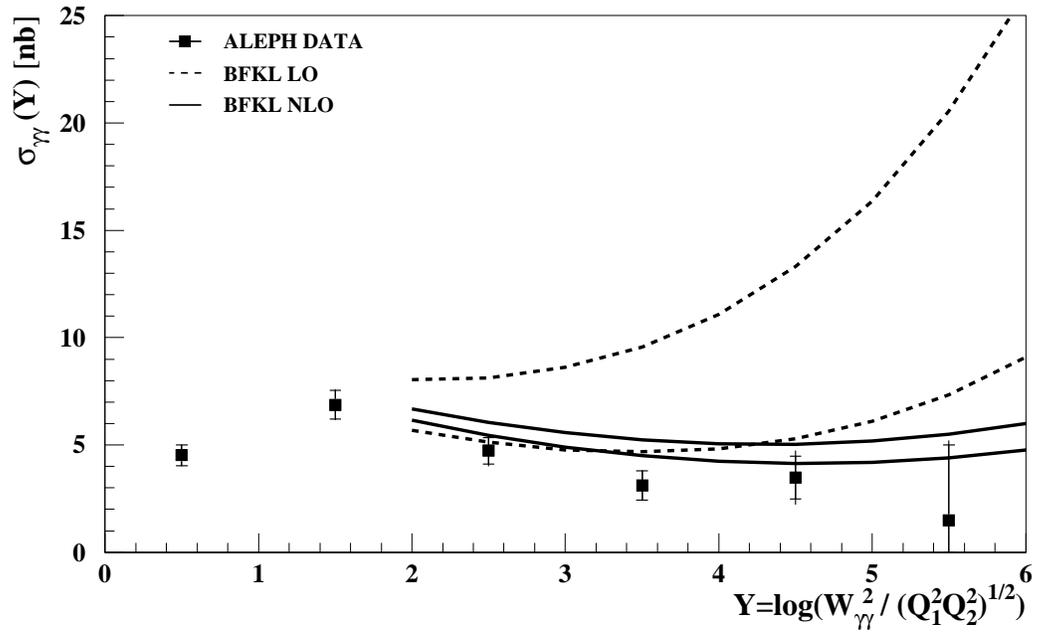}
    \caption{\footnotesize 
    The $\sigma_{\gamma \gamma}$ cross section as a 
 function of $Y$ in comparison with LO BFKL and NLO BFKL, the range 
 giving the uncertainty estimated 
 from the variation of the Regge scale parameter 
 from $Q^2$ to $10Q^2$ and from $Q^2$ to $4Q^2$, respectively.}
    \label{fig:Y_bfkl}
  \end{center}
\end{figure}

The $\sigma_{\gamma \gamma}$ cross 
section~\cite{Budnev:1974de,Nisius:1999cv,Brodsky:1997sd}
was extracted from
the measured $\mathrm{e}^+ \mathrm{e}^-$ cross section using GALUGA 
\cite{Schuler:1997ex}.
The resulting cross section as a
function of $Y$ is plotted in Fig.~\ref{fig:Y_bfkl}, where it is compared 
to LO BFKL and NLO BFKL 
calculations \cite{Kuraev:1977fs,balitski1978,Kim:1999qq,Kim:1999gp,priv:kim}.
In the LO BFKL calculation the Regge scale parameter was 
varied from $Q^2$ to $10Q^2$. Even at the lowest $Y$ values this 
calculation is barely in agreement with the data. 
The NLO BFKL calculation, however, with
the Regge scale parameter varied from $Q^2$ to $4Q^2$ (solid curves) 
is consistent with the data.

\section{Conclusions}

The cross section for the process
\begin{equation}
 \mathrm{e}^+ \mathrm{e}^- \rightarrow 
\mathrm{e}^+ \mathrm{e}^- \gamma \gamma
\rightarrow \mathrm{e}^+ \mathrm{e}^- \mathrm{hadrons}
  \label{eq:ggee}
\end{equation}
has been measured using ALEPH data taken at $\mathrm{e}^+ \mathrm{e}^- $ 
centre-of-mass energies $\sqrt{s} = 
188 - 209\,\mathrm{GeV}$ and 
corresponding to an integrated luminosity of 
$640\,\mathrm{pb}^{-1}$.
The phase space is defined by the electron energies $E_{1,2}' >
0.3 E_\mathrm{beam}$, the polar angles of the electrons 
$0.35\,\mathrm{mrad} < \theta_{1,2} < 155\,\mathrm{mrad}$, and
the mass of the hadronic system $W_{\gamma\gamma} > 3\,\mathrm{GeV}/c^2$.

The differential cross section for 
$\mathrm{e}^+ \mathrm{e}^- \rightarrow \mathrm{e}^+ \mathrm{e}^- \gamma \gamma \rightarrow \mathrm{e}^+ \mathrm{e}^- \mathrm{X}$,
X being the hadronic final state, has been measured as a
function of various event observables.

The majority of measured distributions are well described in shape 
by the PHOT02 and PYTHIA Monte Carlo models, but both required an 
adjustment to their normalization to match the data.
The cross section of the PYTHIA Monte Carlo generator was reduced by 
12\% and the cross section extracted from PHOT02 was increased by 30\%. 

The NLO QCD prediction yields a slightly low cross section.
With the exception of the deep inelastic scattering variable $x_i$
this calculation also gives a reasonable description of the data.
Similar conclusions have been drawn by the L3 \cite{l3_double:1999}
and the OPAL \cite{opal_double:2000} Collaborations.

A slight
enhancement of the data with respect to the simulations
is observed
at high $Y$, but the steep rise of the cross section
as predicted by LO BFKL is not observed in data. However, NLO BFKL is 
close in shape and normalization to the data.

\subsection*{Acknowledgements}

We wish to thank our colleagues in the CERN accelerator divisions for
the successful operation of LEP. We are indebted to the engineers and
technicians in all our institutions for their contribution to the
excellent performance of ALEPH. Those of us from non-member
countries thank CERN for its hospitality.

\newpage
\clearpage

\newpage
\clearpage

\clearpage

\newpage

\begin{figure}[!htp]
  \begin{center}
    \includegraphics[width=0.85\textwidth]{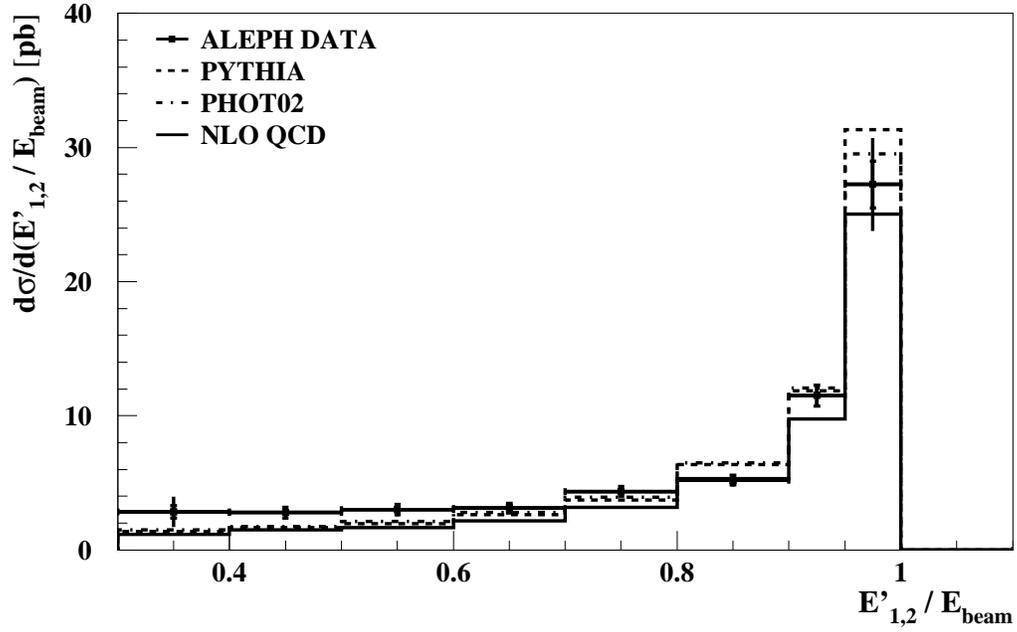}
    \caption{\footnotesize
    Differential cross section as a function of the 
      energy of the scattered electrons 
    $E_\mathrm{1,2} / E_\mathrm{beam}$}
    \label{fig:etag}
  \end{center}
\end{figure}

\begin{figure}[!htp]
  \begin{center}
    \includegraphics[width=0.85\textwidth]{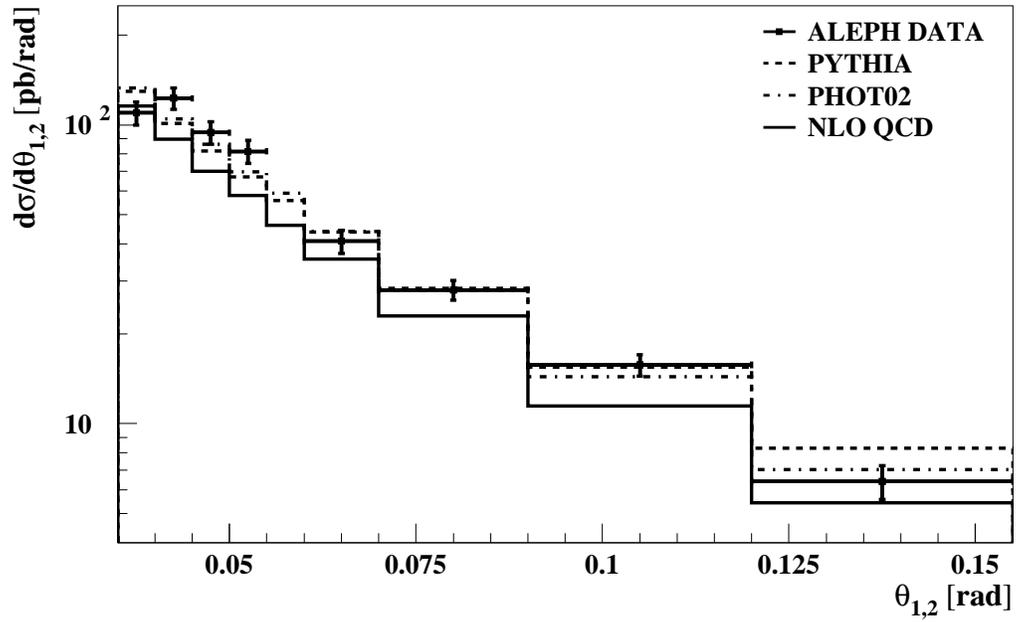}
    \caption{\footnotesize
    Differential cross section as a function of the 
      polar angle $\theta$ of the scattered electrons 
      }
    \label{fig:theta_tag}
  \end{center}
\end{figure}

\begin{figure}[!htp]
  \begin{center}
    \includegraphics[width=0.85\textwidth]{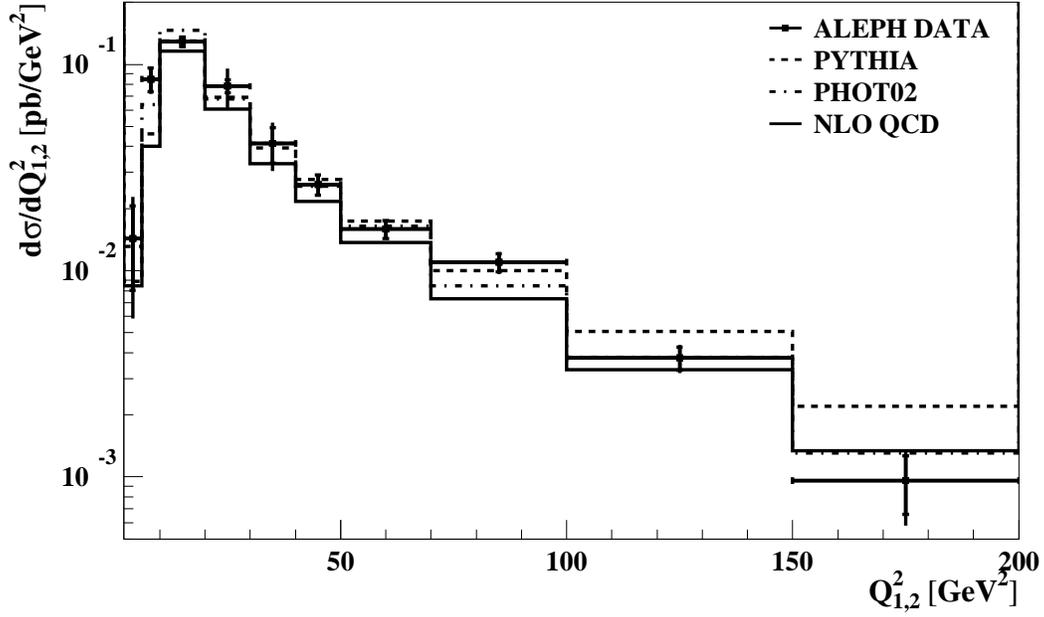}
    \caption{\footnotesize
    Differential cross section as a function of
      the virtualities $Q^2_{1,2}$ of the two photons}
    \label{fig:q2_tag}
  \end{center}
\end{figure}  

\begin{figure}[!htp]
  \begin{center}
    \includegraphics[width=0.85\textwidth]{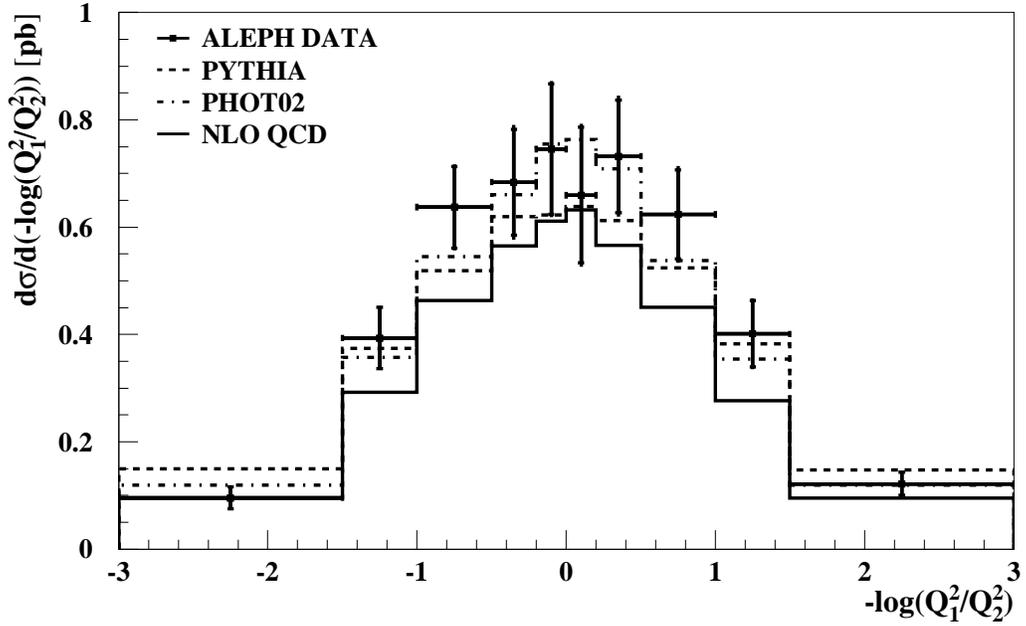}
    \caption{\footnotesize
      Differential cross section as a function of
      the ratio of the two virtualities, shown as $\Delta Q = -\log Q^2_1 /
      Q^2_2$
      }
    \label{fig:q1q2_tag}
  \end{center}
\end{figure}  

\begin{figure}[!htp]
  \begin{center}
    \includegraphics[width=0.85\textwidth]{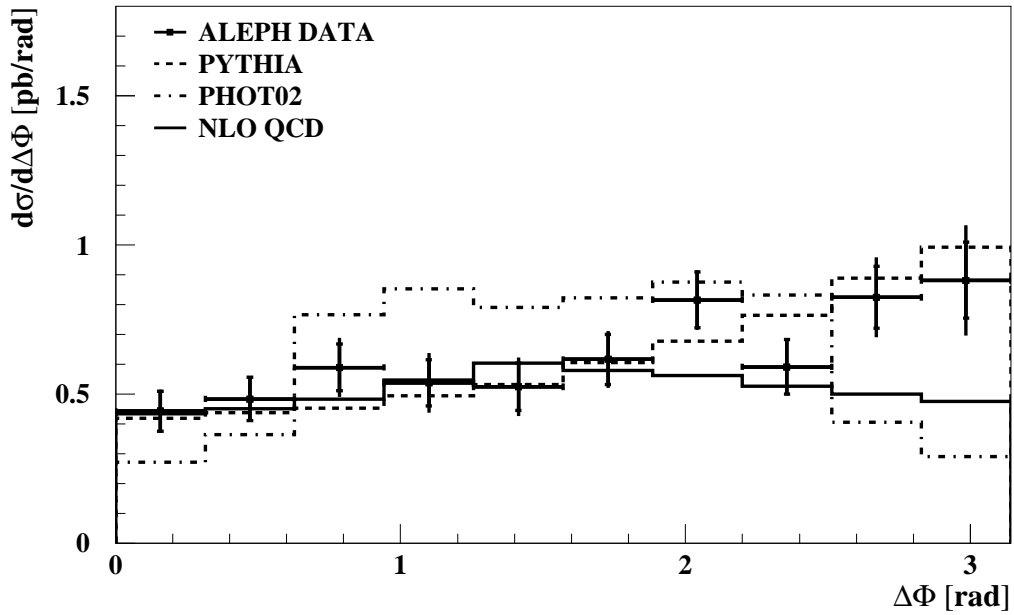}
    \caption{\footnotesize 
      Differential cross section as a function of the
      acoplanarity angle $\Delta\phi$ between the scattered electrons.
      }
    \label{fig:DPHI_tag}
  \end{center}
\end{figure}

\begin{figure}[!htp]
  \begin{center}
    \includegraphics[width=0.85\textwidth]{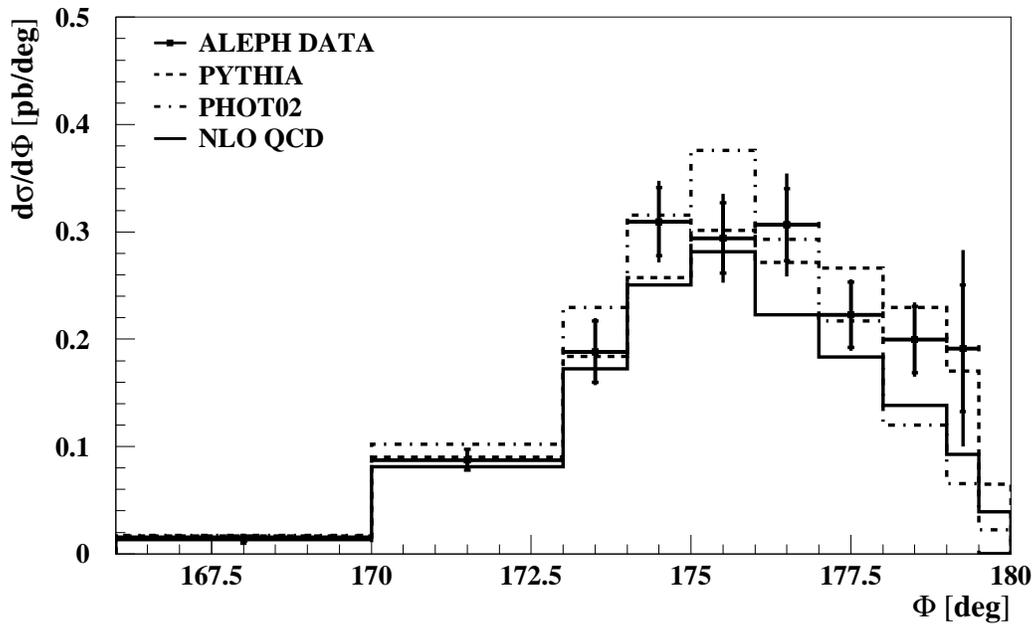}
    \caption{\footnotesize 
    Differential cross section as a function of the
      acolinearity $\Phi$ between the scattered electrons.
      }
    \label{fig:Phi_tag}
  \end{center}
\end{figure}
\clearpage

\begin{figure}[!htp]
  \begin{center}
    \includegraphics[width=0.85\textwidth]{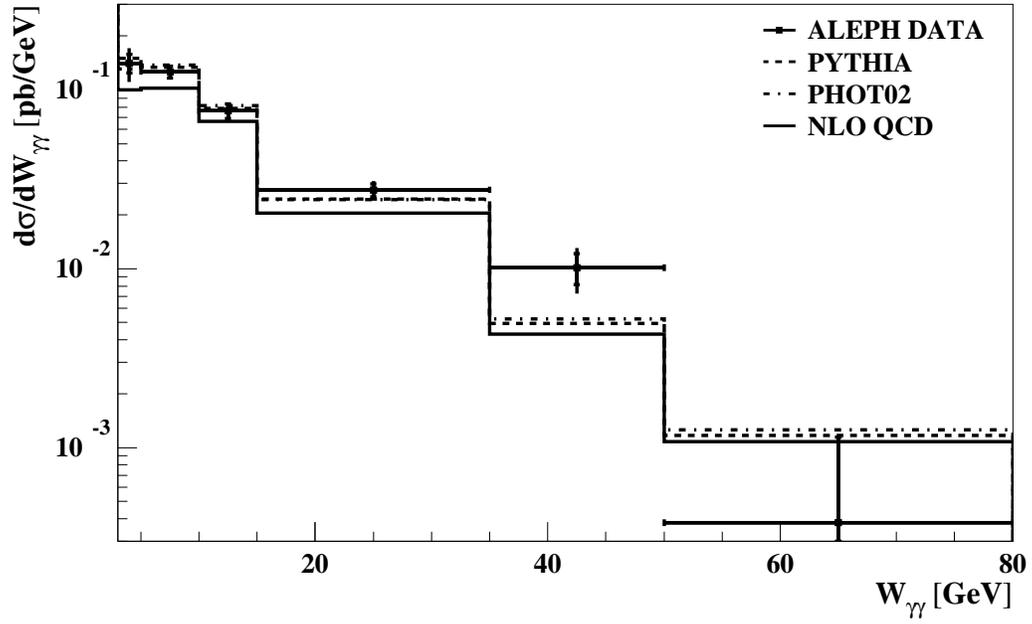}
    \caption{\footnotesize 
    Differential cross section as a function of the mass 
      of the hadronic system $W_{\gamma\gamma}$}
    \label{fig:Wgg}
  \end{center}
\end{figure}

\begin{figure}[!htp]
  \begin{center}
    \includegraphics[width=0.85\textwidth]{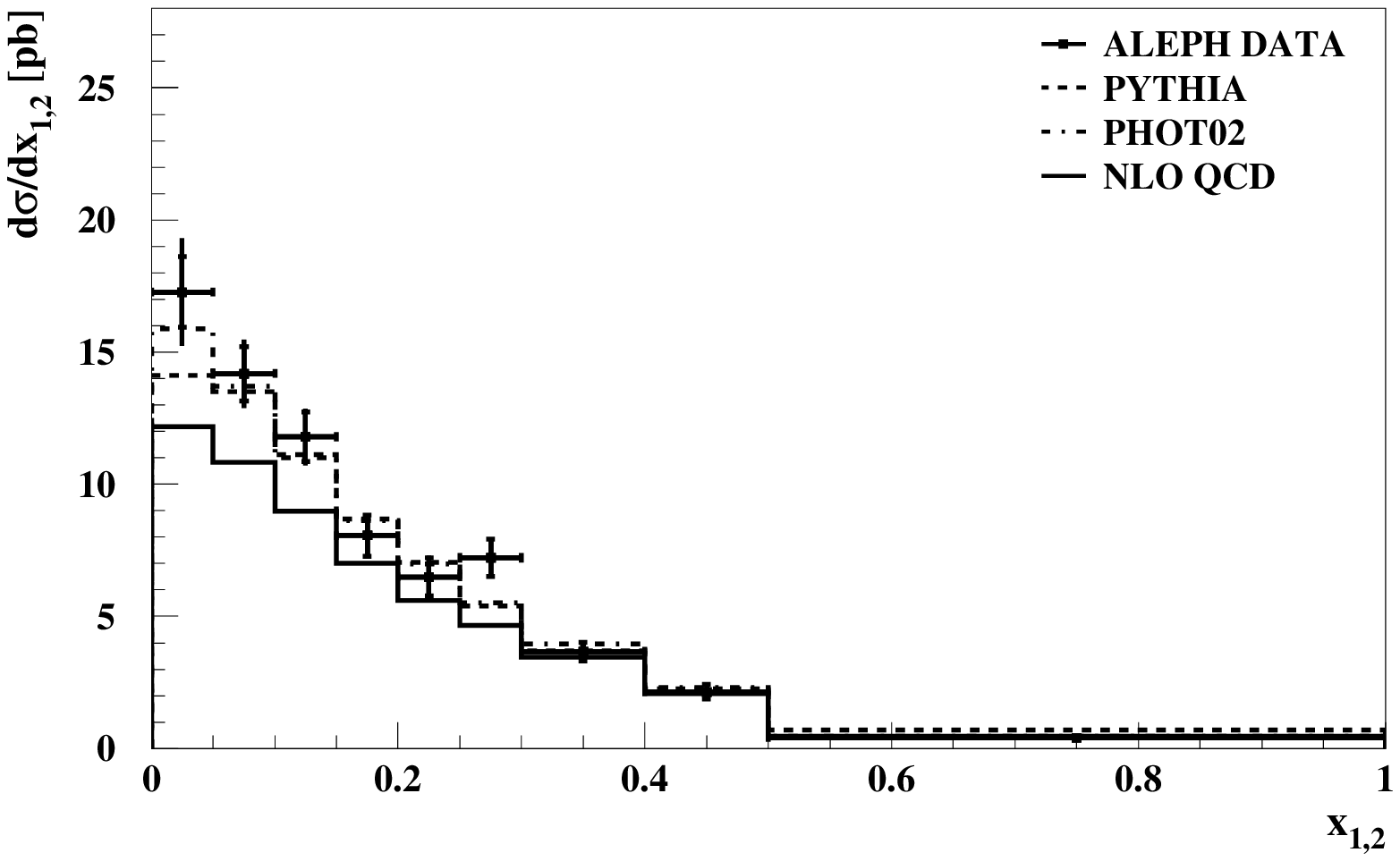}
    \caption{\footnotesize 
    Differential cross section as a function 
      of the deep inelastic scattering variable $x_i$}
    \label{fig:x}
  \end{center}
\end{figure}

\begin{figure}[!htp]
  \begin{center}
    \includegraphics[width=0.85\textwidth]{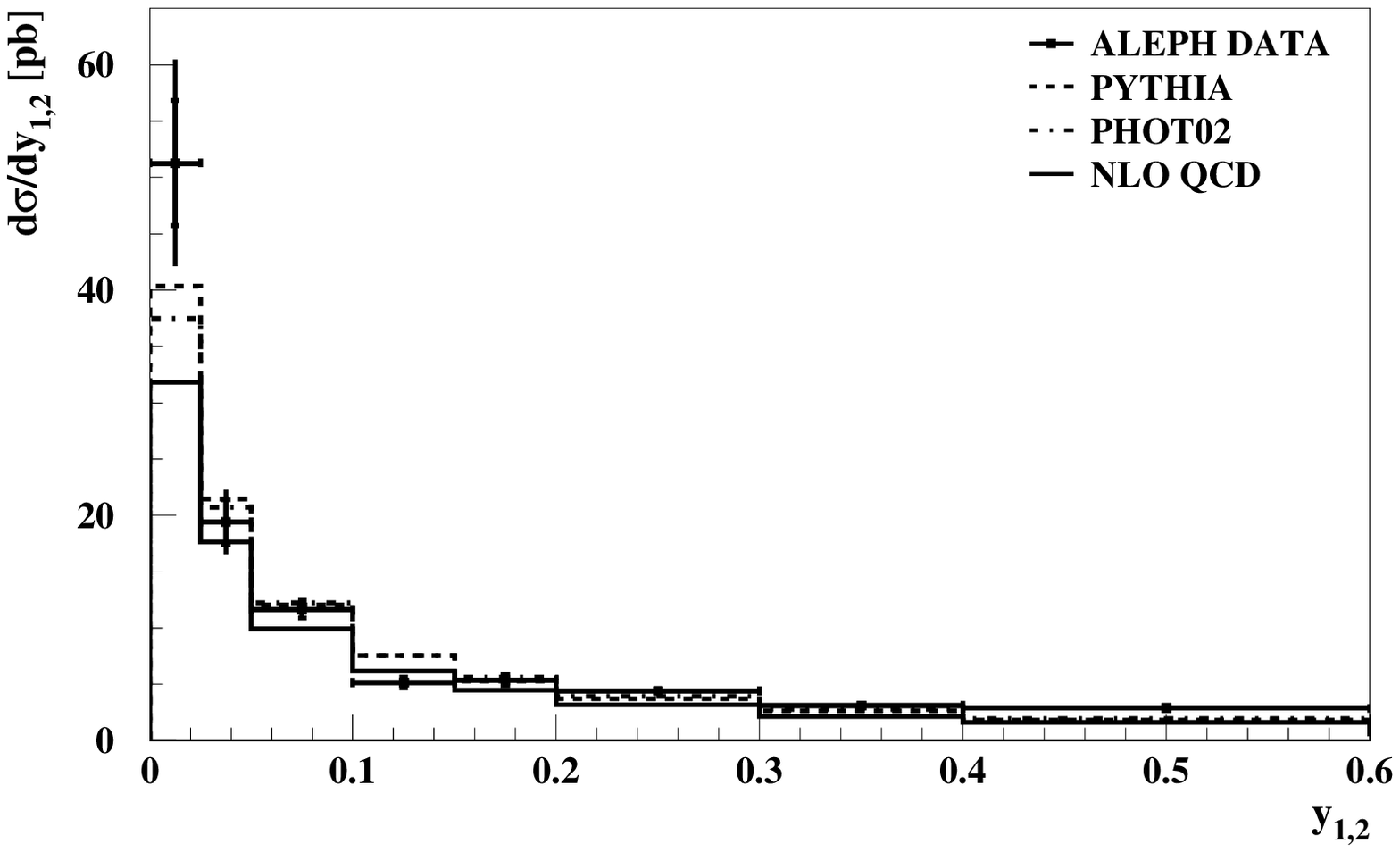}
    \caption{\footnotesize 
    Differential cross section as a function 
      of the deep inelastic scattering variable $y_i$}
    \label{fig:y}
  \end{center}
\end{figure}

\begin{figure}[!htp]
  \begin{center}
    \includegraphics[width=0.85\textwidth]{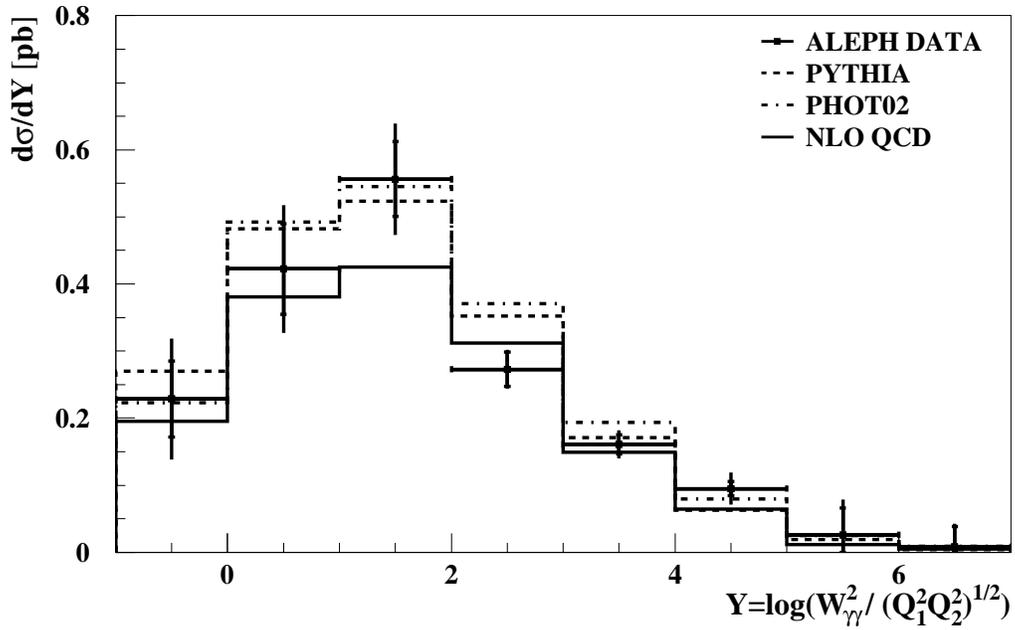}
    \caption{\footnotesize 
    Differential cross section as a function of Y}
    \label{fig:Y}
  \end{center}
\end{figure}

\clearpage

\noindent
{\bf \Large Appendix}

{\small

\begin{table}[!htp]
  \begin{center}
    \caption{\footnotesize 
    Differential cross section as a function of the relative energy 
      of the electrons 
      $e_\mathrm{1,2} = \frac{E_\mathrm{1,2}'}{E_\mathrm{beam}}$.}
\vspace*{3mm}
    \begin{tabular}{c @{ -- } c | c @{ $\pm$ } c @{ $\pm$ } c c c c c }
      \multicolumn{2}{c|}{$e_\mathrm{1,2}$~bin } &
      \multicolumn{6}{c}{\raisebox{1.0mm}[0mm][0mm]
      {$d\sigma / de_\mathrm{1,2} \quad [\mathrm{pb}]$ }} \\

      \multicolumn{2}{c|}{~} &
      \multicolumn{3}{c}{data}
      & PYTHIA &
      PHOT02 & QCD \\ \hline
  0.30 &  0.40 &  2.85 &  0.47 &  1.03 &  1.37 &   1.51 &  1.18 \\ 
  0.40 &  0.50 &  2.78 &  0.39 &  0.30 &  1.69 &   1.74 &  1.50 \\ 
  0.50 &  0.60 &  3.01 &  0.37 &  0.32 &  1.98 &   2.14 &  1.69 \\ 
  0.60 &  0.70 &  3.12 &  0.36 &  0.10 &  2.63 &   2.78 &  2.18 \\ 
  0.70 &  0.80 &  4.36 &  0.36 &  0.21 &  3.72 &   3.94 &  3.20 \\ 
  0.80 &  0.90 &  5.19 &  0.36 &  0.27 &  6.35 &   6.51 &  5.31 \\ 
  0.90 &  0.95 & 11.51 &  0.78 &  0.40 & 11.86 &  12.06 &  9.76 \\ 
  0.95 &  1.00 & 27.24 &  1.75 &  2.99 & 31.33 &  29.54 & 25.02 \\ \hline
\multicolumn{7}{c}{}\\
\multicolumn{2}{c}{} &\multicolumn{3}{c}{
  \raisebox{2.0mm}[0mm][0mm]{$\chi^2 / $ NDF}} &
\raisebox{2.0mm}[0mm][0mm]{23.3 / 7} &    
\raisebox{2.0mm}[0mm][0mm]{20.2 / 7}  &    
\raisebox{2.0mm}[0mm][0mm]{34.6 / 8} \\ \hline
    \end{tabular}
    \label{tab:e_tag}
  \end{center}
\end{table}

\begin{table}[!htp]
  \begin{center}
    \caption{\footnotesize 
    Differential cross section as a function of the
      polar angle $\theta$ of the scattered electrons.}
\vspace*{3mm}
    \begin{tabular}{c @{ -- } c | c @{ $\pm$ } c @{ $\pm$ } c c c c c }
      \multicolumn{2}{c|}{$\theta$~bin } &
      \multicolumn{6}{c}{\raisebox{1.0mm}[0mm][0mm]
      {$d\sigma / d\theta \quad [\mathrm{pb/rad}]$ }} \\

      \multicolumn{2}{c|}{[rad]} &
      \multicolumn{3}{c}{data}
      & PYTHIA &
      PHOT02 & QCD \\ \hline

    0.035 &    0.040 &  110.11 &    9.63 &    3.92 &  129.33 &   132.93 &  115.79 \\ 
    0.040 &    0.045 &  122.18 &    9.94 &    3.01 &  101.10 &   104.87 &   89.68 \\ 
    0.045 &    0.050 &   94.71 &    8.01 &    2.47 &   81.76 &    86.08 &   69.86 \\ 
    0.050 &    0.055 &   81.16 &    7.08 &    2.73 &   66.91 &    69.70 &   57.97 \\ 
    0.060 &    0.070 &   40.57 &    3.63 &    1.65 &   43.84 &    44.06 &   35.49 \\ 
    0.070 &    0.090 &   28.08 &    2.14 &    0.92 &   28.37 &    28.03 &   22.92 \\ 
    0.090 &    0.120 &   15.81 &    1.29 &    0.55 &   15.46 &    14.38 &   11.44 \\ 
    0.120 &    0.155 &    6.49 &    0.84 &    0.27 &    8.30 &     7.03 &    5.44 \\  \hline
\multicolumn{7}{c}{}\\
\multicolumn{2}{c}{} &\multicolumn{3}{c}{
  \raisebox{2.0mm}[0mm][0mm]{$\chi^2 / $ NDF}} &
\raisebox{2.0mm}[0mm][0mm]{18.4 / 7} &    
\raisebox{2.0mm}[0mm][0mm]{13.1 / 7}  &    
\raisebox{2.0mm}[0mm][0mm]{45.8 / 8} \\ \hline
       \end{tabular}
    \label{tab:theta_tag}
  \end{center}
\end{table}

\begin{table}[!htp]
  \begin{center}{\small
    \caption{\footnotesize 
    Differential cross section as a function of the 
virtuality $Q^2_{1,2}$ of the photons.}
\vspace*{3mm}
    \begin{tabular}{c @{ -- } c | c @{ $\pm$ } c @{ $\pm$ } c c c c c }
      \multicolumn{2}{c|}{$Q^2$~bin } &
      \multicolumn{6}{c}{\raisebox{1.0mm}[0mm][0mm]
      {$d\sigma / dQ^2_i \quad [\mathrm{pb/GeV^2}]$ }} \\

      \multicolumn{2}{c|}{[GeV]} &
      \multicolumn{3}{c}{data}
      & PYTHIA &
      PHOT02 & QCD \\ \hline

   2.0 &   6.0 &   0.0133 &   0.0062 &   0.0056 &   0.0089 &   0.0131 &   0.0085 \\ 
   6.0 &  10.0 &   0.0852 &   0.0112 &   0.0059 &   0.0461 &   0.0638 &   0.0401 \\ 
  10.0 &  20.0 &   0.1290 &   0.0070 &   0.0052 &   0.1295 &   0.1462 &   0.1160 \\ 
  20.0 &  30.0 &   0.0785 &   0.0056 &   0.0161 &   0.0691 &   0.0682 &   0.0606 \\ 
  30.0 &  40.0 &   0.0415 &   0.0081 &   0.0073 &   0.0414 &   0.0393 &   0.0331 \\ 
  40.0 &  50.0 &   0.0263 &   0.0030 &   0.0013 &   0.0278 &   0.0256 &   0.0216 \\ 
  50.0 &  70.0 &   0.0160 &   0.0016 &   0.0004 &   0.0174 &   0.0165 &   0.0137 \\ 
  70.0 & 100.0 &   0.0111 &   0.0011 &   0.0008 &   0.0100 &   0.0084 &   0.0073 \\ 
 100.0 & 150.0 &   0.0038 &   0.0005 &   0.0003 &   0.0051 &   0.0038 &   0.0033 \\ 
 150.0 & 200.0 &   0.0010 &   0.0003 &   0.0002 &   0.0022 &   0.0013 &   0.0013 \\ \hline
\multicolumn{7}{c}{}\\
\multicolumn{2}{c}{} &\multicolumn{3}{c}{
  \raisebox{2.0mm}[0mm][0mm]{$\chi^2 / $ NDF}} &
\raisebox{2.0mm}[0mm][0mm]{27.0 / 9} &    
\raisebox{2.0mm}[0mm][0mm]{12.0 / 9}  &    
\raisebox{2.0mm}[0mm][0mm]{30.3 / 10} \\ \hline
    \end{tabular}}
    \label{tab:q2_tag}
  \end{center}
\end{table}

\begin{table}[!htp]
  \begin{center}{\small
    \caption{\footnotesize 
    Differential cross section as a function of the ratio of the 
virtualities of the two photons $\Delta Q = -\log \frac{Q_1^2}{Q_2^2}$.}
\vspace*{3mm}
    \begin{tabular}{c @{ -- } c | c @{ $\pm$ } c @{ $\pm$ } c c c c c }
      \multicolumn{2}{c|}{$\Delta Q$~bin } &
      \multicolumn{6}{c}{\raisebox{1.0mm}[0mm][0mm]
      {$d\sigma / d\Delta Q \quad [\mathrm{pb}]$ }} \\

      \multicolumn{2}{c|}{[GeV]} &
      \multicolumn{3}{c}{data}
      & PYTHIA &
      PHOT02 & QCD \\ \hline
 $-$3.0 & $-$1.5 &  0.096 &  0.021 &  0.007 &  0.149 &  0.120 &  0.096 \\ 
 $-$1.5 & $-$1.0 &  0.394 &  0.057 &  0.016 &  0.374 &  0.357 &  0.292 \\ 
 $-$1.0 & $-$0.5 &  0.631 &  0.076 &  0.025 &  0.519 &  0.546 &  0.464 \\ 
 $-$0.5 & $-$0.2 &  0.686 &  0.098 &  0.042 &  0.619 &  0.661 &  0.565 \\ 
 $-$0.2 &  0.0 &  0.749 &  0.122 &  0.034 &  0.623 &  0.755 &  0.611 \\ 
  0.0 &  0.2 &  0.664 &  0.127 &  0.044 &  0.638 &  0.763 &  0.633 \\ 
  0.2 &  0.5 &  0.736 &  0.105 &  0.043 &  0.612 &  0.709 &  0.566 \\ 
  0.5 &  1.0 &  0.620 &  0.083 &  0.035 &  0.524 &  0.538 &  0.451 \\ 
  1.0 &  1.5 &  0.403 &  0.062 &  0.023 &  0.383 &  0.354 &  0.276 \\ 
  1.5 &  3.0 &  0.122 &  0.021 &  0.005 &  0.147 &  0.120 &  0.095 \\ 
 \hline
\multicolumn{7}{c}{}\\
\multicolumn{2}{c}{} &\multicolumn{3}{c}{
  \raisebox{2.0mm}[0mm][0mm]{$\chi^2 / $ NDF}} &
\raisebox{2.0mm}[0mm][0mm]{13.1 / 9} &    
\raisebox{2.0mm}[0mm][0mm]{ 4.7 / 9}  &    
\raisebox{2.0mm}[0mm][0mm]{20.9 /  10} \\ \hline
    \end{tabular}}
    \label{tab:q1q2_tag}
  \end{center}
\end{table}

\begin{table}[!htp]
  \begin{center}
    \caption{\footnotesize 
    Differential cross section as a function of the acolinearity 
      $\Phi$ of the scattered electron and the scattered positron.}
{\small
\vspace*{3mm}
    \begin{tabular}{c @{ -- } c | c @{ $\pm$ } c @{ $\pm$ } c c c c c }
      \multicolumn{2}{c|}{$\Phi$~bin } &
      \multicolumn{6}{c}{\raisebox{1.0mm}[0mm][0mm]
      {$d\sigma / d\Phi \quad [\mathrm{pb/degree}]$ }} \\
      \multicolumn{2}{c|}{[degree]} &
      \multicolumn{3}{c}{data}
      & PYTHIA &
      PHOT02 & QCD \\ \hline
 166.0 & 170.0 &   0.014 &   0.004 &   0.001 &   0.017 &     0.017 &   0.015 \\ 
 170.0 & 173.0 &   0.088 &   0.010 &   0.002 &   0.090 &     0.102 &   0.081 \\ 
 173.0 & 174.0 &   0.189 &   0.029 &   0.013 &   0.184 &     0.230 &   0.172 \\ 
 174.0 & 175.0 &   0.307 &   0.032 &   0.021 &   0.257 &     0.316 &   0.251 \\ 
 175.0 & 176.0 &   0.295 &   0.033 &   0.025 &   0.301 &     0.376 &   0.281 \\ 
 176.0 & 177.0 &   0.308 &   0.034 &   0.034 &   0.271 &     0.293 &   0.223 \\ 
 177.0 & 178.0 &   0.220 &   0.030 &   0.014 &   0.266 &     0.217 &   0.184 \\ 
 178.0 & 179.0 &   0.200 &   0.031 &   0.016 &   0.229 &     0.120 &   0.138 \\ 
 179.0 & 179.5 &   0.192 &   0.059 &   0.070 &   0.170 &     0.066 &   0.093 \\ \hline
\multicolumn{7}{c}{}\\
\multicolumn{2}{c}{} &\multicolumn{3}{c}{
  \raisebox{2.0mm}[0mm][0mm]{$\chi^2 / $ NDF}} &
\raisebox{2.0mm}[0mm][0mm]{ 5.9 / 8} &    
\raisebox{2.0mm}[0mm][0mm]{15.6 / 8}  &    
\raisebox{2.0mm}[0mm][0mm]{12.1 /  9} \\ \hline
    \end{tabular}}
    \label{tab:bhabha}
  \end{center}
\end{table}

\begin{table}[!htp]
  \begin{center}
    \caption{\footnotesize 
    Differential cross section as a function of the acoplanarity 
      angle 
      $\Delta\Phi$ of the scattered electron and the scattered positron.}
{\small
\vspace*{3mm}
    \begin{tabular}{c @{ -- } c | c @{ $\pm$ } c @{ $\pm$ } c c c c c }
      \multicolumn{2}{c|}{$\Delta\phi$~bin } &
      \multicolumn{6}{c}{\raisebox{1.0mm}[0mm][0mm]
      {$d\sigma / d\Delta\phi \quad [\mathrm{pb/degree}]$ }} \\

      \multicolumn{2}{c|}{[degree]} &
      \multicolumn{3}{c}{data}
      & PYTHIA &
      PHOT02 & QCD \\ \hline
  0.00 &  0.31 &  0.45 &  0.07 &  0.02 &  0.42 &   0.27 &  0.43 \\ 
  0.31 &  0.63 &  0.49 &  0.07 &  0.02 &  0.44 &   0.36 &  0.45 \\ 
  0.63 &  0.94 &  0.58 &  0.08 &  0.06 &  0.45 &   0.77 &  0.48 \\ 
  0.94 &  1.26 &  0.54 &  0.08 &  0.07 &  0.49 &   0.85 &  0.55 \\ 
  1.26 &  1.57 &  0.53 &  0.08 &  0.06 &  0.53 &   0.79 &  0.60 \\ 
  1.57 &  1.88 &  0.62 &  0.08 &  0.05 &  0.61 &   0.82 &  0.58 \\ 
  1.88 &  2.20 &  0.82 &  0.09 &  0.03 &  0.68 &   0.88 &  0.56 \\ 
  2.20 &  2.51 &  0.58 &  0.09 &  0.03 &  0.76 &   0.83 &  0.53 \\ 
  2.51 &  2.83 &  0.83 &  0.10 &  0.08 &  0.89 &   0.41 &  0.50 \\ 
  2.83 & $\pi$ &  0.88 &  0.13 &  0.14 &  0.99 &   0.29 &  0.48 \\ \hline
\multicolumn{7}{c}{}\\
\multicolumn{2}{c}{} &\multicolumn{3}{c}{
  \raisebox{2.0mm}[0mm][0mm]{$\chi^2 / $ NDF}} &
\raisebox{2.0mm}[0mm][0mm]{ 8.7 / 9} &    
\raisebox{2.0mm}[0mm][0mm]{60.7 / 9}  &    
\raisebox{2.0mm}[0mm][0mm]{19.5 / 10} \\ \hline
    \end{tabular}}
    \label{tab:DPHI_tag}
  \end{center}
\end{table}

\begin{table}[!htp]
  \begin{center}
    \caption{\footnotesize 
    Differential cross section as a function of the mass
      $W_{\gamma \gamma}$ of the hadronic system.}
{\small
\vspace*{3mm}
    \begin{tabular}{c @{ -- } c | c @{ $\pm$ } c @{ $\pm$ } c c c c c }
      \multicolumn{2}{c|}{$W_{\gamma \gamma}$~bin } &
      \multicolumn{6}{c}{\raisebox{1.0mm}[0mm][0mm]
      {$d\sigma / dW_{\gamma \gamma} \quad [\mathrm{pb/GeV}]$ }} \\

      \multicolumn{2}{c|}{[GeV]} &
      \multicolumn{3}{c}{data}
      & PYTHIA &
      PHOT02 & QCD \\ \hline
   3.0 &   5.0 &   0.1409 &   0.0164 &   0.0257 &   0.1499 &   0.1307 &   0.0996 \\ 
   5.0 &  10.0 &   0.1263 &   0.0095 &   0.0033 &   0.1334 &   0.1380 &   0.1019 \\ 
  10.0 &  15.0 &   0.0763 &   0.0072 &   0.0032 &   0.0787 &   0.0814 &   0.0665 \\ 
  15.0 &  35.0 &   0.0276 &   0.0023 &   0.0023 &   0.0246 &   0.0244 &   0.0205 \\ 
  35.0 &  50.0 &   0.0102 &   0.0020 &   0.0021 &   0.0050 &   0.0052 &   0.0043 \\ 
  50.0 &  80.0 &   
\multicolumn{1}{c}{\hspace*{-10pt}0.0005} 
  & $^{+0.0010}_{-0.0005}$ & 0.0020
&   0.0012 &     0.0013 &   0.0011 \\ \hline
\multicolumn{8}{c}{}\\
\multicolumn{2}{c}{} &\multicolumn{3}{c}{
  \raisebox{2.0mm}[0mm][0mm]{$\chi^2 / $ NDF}} &
\raisebox{2.0mm}[0mm][0mm]{ 5.3 / 5} &    
\raisebox{2.0mm}[0mm][0mm]{ 6.4 / 5}  &    
\raisebox{2.0mm}[0mm][0mm]{18.0 /  6} \\ \hline
    \end{tabular}}
    \label{tab:Wgg}
  \end{center}
\end{table}

\begin{table}[!htp]
  \begin{center}
    \caption{\footnotesize 
    Differential cross section as a function of
      the Bj\"orken variable $x_{1,2}$.}
{\small
\vspace*{3mm}
    \begin{tabular}{c @{ -- } c | c @{ $\pm$ } c @{ $\pm$ } c c c c c }
      \multicolumn{2}{c|}{$x_i$~bin } &
      \multicolumn{6}{c}{\raisebox{1.0mm}[0mm][0mm]
      {$d\sigma / dx_i \quad [\mathrm{pb}]$ }} \\

      \multicolumn{2}{c|}{~} &
      \multicolumn{3}{c}{data}
      & PYTHIA &
      PHOT02 & QCD \\ \hline
  0.00 &  0.05 & 17.71 &  1.38 &  2.00 & 14.11 &  15.88 & 12.17 \\ 
  0.05 &  0.10 & 14.48 &  1.05 &  1.13 & 13.51 &  13.70 & 10.84 \\ 
  0.10 &  0.15 & 11.97 &  0.95 &  0.77 & 11.12 &  11.00 &  8.99 \\ 
  0.15 &  0.20 &  8.20 &  0.79 &  0.44 &  8.67 &   8.62 &  7.02 \\ 
  0.20 &  0.25 &  6.62 &  0.72 &  0.56 &  7.03 &   6.98 &  5.61 \\ 
  0.25 &  0.30 &  7.31 &  0.70 &  0.43 &  5.40 &   5.50 &  4.66 \\ 
  0.30 &  0.40 &  3.71 &  0.36 &  0.22 &  3.68 &   3.96 &  3.45 \\ 
  0.40 &  0.50 &  2.19 &  0.28 &  0.14 &  2.25 &   2.31 &  2.07 \\ 
  0.50 &  1.00 &  0.42 &  0.06 &  0.09 &  0.70 &   0.46 &  0.46 \\ \hline
\multicolumn{7}{c}{}\\
\multicolumn{2}{c}{} &\multicolumn{3}{c}{
  \raisebox{2.0mm}[0mm][0mm]{$\chi^2 / $ NDF}} &
\raisebox{2.0mm}[0mm][0mm]{16.0 / 8} &    
\raisebox{2.0mm}[0mm][0mm]{7.3 / 8}  &    
\raisebox{2.0mm}[0mm][0mm]{30.6 / 9} \\ \hline
    \end{tabular}}
    \label{tab:xi_tag}
  \end{center}
\end{table}

\begin{table}[!htp]
  \begin{center}
    \caption{\footnotesize 
    Differential cross section as a function of
      the Bj\"orken variable $y_{1,2}$.}
{\small
\vspace*{3mm}
    \begin{tabular}{c @{ -- } c | c @{ $\pm$ } c @{ $\pm$ } c c c c c }
      \multicolumn{2}{c|}{$y$~bin } &
      \multicolumn{6}{c}{\raisebox{1.0mm}[0mm][0mm]
      {$d\sigma / dy_i \quad [\mathrm{pb}]$ }} \\

      \multicolumn{2}{c|}{~} &
      \multicolumn{3}{c}{data}
      & PYTHIA &
      PHOT02 & QCD \\ \hline
  0.000 &  0.025 & 51.28 &  5.54 &  7.34 & 40.37 &  37.49 & 31.79 \\ 
  0.025 &  0.050 & 19.41 &  1.99 &  2.07 & 21.48 &  20.73 & 17.61 \\ 
  0.050 &  0.100 & 11.68 &  0.79 &  0.49 & 12.02 &  12.27 &  9.93 \\ 
  0.100 &  0.150 &  5.16 &  0.50 &  0.35 &  7.54 &   7.55 &  6.19 \\ 
  0.150 &  0.200 &  5.39 &  0.52 &  0.21 &  5.31 &   5.61 &  4.49 \\ 
  0.200 &  0.300 &  4.43 &  0.36 &  0.22 &  3.74 &   3.94 &  3.21 \\ 
  0.300 &  0.400 &  3.15 &  0.36 &  0.09 &  2.64 &   2.79 &  2.20 \\ 
  0.400 &  0.600 &  2.85 &  0.27 &  0.26 &  1.84 &   1.95 &  1.60 \\ 
  0.600 &  0.800 &  1.41 &  0.23 &  0.50 &  0.69 &   0.76 &  0.59 \\ \hline
\multicolumn{7}{c}{}\\
\multicolumn{2}{c}{} &\multicolumn{3}{c}{
  \raisebox{2.0mm}[0mm][0mm]{$\chi^2 / $ NDF}} &
\raisebox{2.0mm}[0mm][0mm]{31.1 / 8} &    
\raisebox{2.0mm}[0mm][0mm]{28.1 / 8}  &    
\raisebox{2.0mm}[0mm][0mm]{42.4 /  9} \\ \hline
    \end{tabular}}
    \label{tab:yi_tag}
  \end{center}
\end{table}

\begin{table}[!htp]
  \begin{center}
    \caption{\footnotesize 
    Differential cross section as a function of $Y$.}
{\small
\vspace*{3mm}
    \begin{tabular}{c @{ -- } c | c @{ $\pm$ } c @{ $\pm$ } c c c c c }
      \multicolumn{2}{c|}{$Y$~bin } &
      \multicolumn{6}{c}{\raisebox{1.0mm}[0mm][0mm]
      {$d\sigma / dY \quad [\mathrm{pb}]$ }} \\

      \multicolumn{2}{c|}{~} &
      \multicolumn{3}{c}{data}
      & PYTHIA &
      PHOT02 & QCD \\ \hline
 $-$1.0 &  0.0 &  0.224 &  0.028 &  0.031 &  0.270 &   0.223 &  0.195 \\ 
  0.0 &  1.0 &  0.437 &  0.039 &  0.015 &  0.483 &   0.493 &  0.381 \\ 
  1.0 &  2.0 &  0.542 &  0.045 &  0.022 &  0.523 &   0.545 &  0.425 \\ 
  2.0 &  3.0 &  0.422 &  0.040 &  0.030 &  0.352 &   0.370 &  0.312 \\ 
  3.0 &  4.0 &  0.227 &  0.035 &  0.017 &  0.171 &   0.194 &  0.149 \\ 
  4.0 &  5.0 &  0.117 &  0.026 &  0.024 &  0.062 &   0.079 &  0.065 \\ 
  5.0 &  6.0 &  
\multicolumn{1}{c}{\hspace*{-10pt}0.026} 
  & \hspace*{-10pt}$^{+0.040}_{-0.020}$ & 0.013
&  0.019 &   0.025 &  0.012 \\
  6.0 &  7.0 & 
\multicolumn{1}{c}{\hspace*{-10pt}0.008} 
  & \hspace*{-10pt}$^{+0.030}_{-0.008}$ & 0.005
&  0.004 &   0.009 &  0.004 \\ \hline
\multicolumn{7}{c}{}\\
\multicolumn{2}{c}{} &\multicolumn{3}{c}{
  \raisebox{2.0mm}[0mm][0mm]{$\chi^2 / $ NDF}} &
\raisebox{2.0mm}[0mm][0mm]{ 9.3 / 7} &    
\raisebox{2.0mm}[0mm][0mm]{ 5.2 / 7}  &    
\raisebox{2.0mm}[0mm][0mm]{ 19.3 / 8} \\ \hline
    \end{tabular}}
    \label{tab:Y_tag}
  \end{center}
\end{table}

}


\begin{thebibliography}{10}

\bibitem{Sjostrand:2000wi}
T.~Sj{\"o}strand et~al., 
{\em High-energy-physics event generation with PYTHIA 6.1}, 
Comput. Phys. Commun. {\bf 135} (2001) 238.

\bibitem{phot02pap}
{ALEPH Collaboration}, 
{\em An experimental study of $\gamma\gamma \rightarrow $ hadrons at LEP}, 
Phys. Lett. {\bf B313} (1993) 509.

\bibitem{Cacciari:2000cb}
M.~Cacciari, V.~Del Duca, S.~Frixione and Z.~Trocsanyi, 
{\em QCD radiative corrections to $\gamma^* \gamma^* \rightarrow $ hadrons}, 
JHEP {\bf 02} (2001) 29.

\bibitem{Kuraev:1977fs}
E.A.~Kuraev, L.N.~Lipatov and V.S.~Fadin, 
{\em The Pomeranchuk singularity in nonabelian gauge theories}, 
Sov. Phys. JETP {\bf 45} (1977) 199.

\bibitem{balitski1978}
I.I.~Balitsky and L.N.~Lipatov, 
{\em The Pomeranchuk singularity in quantum chromodynamics}, 
Sov. J. Nucl. Phys. {\bf 28} (1978) 822.

\bibitem{Kim:1999qq}
V.T.~Kim, L.N.~Lipatov and G.B.~Pivovarov, 
{\em The next-to-leading dynamics of the BFKL pomeron}, 
Proceedings of the 
29th International Symposium on Multiparticle Dynamics: 
QCD and Multiparticle Production - ISMD '99, Providence, RI, USA, 
Eds.: I.~Sarcevic and C.-I.~Tan, World Sci., Singapore (2000) 79.

\bibitem{Kim:1999gp}
V.T.~Kim, L.N.~Lipatov and G.B.~Pivovarov, 
{\em The next-to-leading BFKL pomeron with optimal renormalization}, 
Proceedings of the 
International Conference on Elastic and Diffractive Scattering 1999, Protvino, 
Russia, 
Eds.: V.A.~Petrov and A.V.~Prokudin, World Sci., Singapore (2000) 237.

\bibitem{priv:kim}
V.T.~Kim, 
private communication.

\bibitem{Friberg:2000ra}
C.~Friberg and T.~Sj{\"o}strand, 
{\em Total cross sections and event properties from real to virtual
photons}, 
JHEP {\bf 09} (2000) 010.

\bibitem{Friberg:2000nx}
C.~Friberg and T.~Sj{\"o}strand, 
{\em Effects of longitudinal photons}, 
Phys. Lett. {\bf B492} (2000) 123.

\bibitem{phojet}
R.~Engel and J.~Ranft, 
{\em Hadronic photon-photon interactions at high energies}, 
Phys. Rev. {\bf D54} (1996) 4244.

\bibitem{Vermaseren:1983cz}
J.A.M.~Vermaseren, 
{\em Two photon processes at very high energies}, 
Nucl. Phys. {\bf B229} (1983) 347.

\bibitem{Cochard:1980iy}
G.~Cochard and P.~Kessler (eds.), 
{\em Gamma gamma collisions}, proceedings, international workshop, Amiens,
France, April 8-12, 1980, Lecture Notes In Physics 134, 
(Berlin: Springer 1980).

\bibitem{Bhattacharya:1977re}
R.~Bhattacharya, J.~Smith and G.~Grammer, Jr.\, 
{\em Two photon production processes at high energy}, 
Phys. Rev. {\bf D15} (1977) 3267.

\bibitem{Ginzburg:1982bs}
I.F.~Ginzburg and V.G.~Serbo, 
{\em Some comments on the total {$\gamma \gamma$} $\to$ hadron
  cross-section at high energy}, 
Phys. Lett. {\bf B109} (1982) 231.

\bibitem{Bonneau:1973kg}
G.~Bonneau, M.G.~Gourdin and F.~Martin, 
{\em Inelastic lepton anti-lepton scattering and the two photon exchange
  approximation}, 
Nucl. Phys. {\bf B54} (1973) 573.

\bibitem{koralz}
S.~Jadach, B.F.L.~Ward and Z.~Was, 
{\em The Monte Carlo program KORALZ, version 4.0, for lepton or quark 
pair production at LEP / SLC energies}, 
Comput. Phys. Commun. {\bf 79} (1994) 503.

\bibitem{Marchesini:1992ch}
G.~Marchesini et~al., 
{\em HERWIG: A Monte Carlo event generator for simulating hadron emission
  reactions with interfering gluons}, 
Comput. Phys. Commun. {\bf 67} (1992) 465.

\bibitem{ALEPHdetperfvdet}
ALEPH Collaboration, 
{\em ALEPH: A detector for electron-positron annihilations at LEP}, 
Nucl. Instrum. and Meth. {\bf A294} (1990) 121; 
ALEPH Collaboration, 
{\em Performance of the ALEPH detector at LEP}, 
Nucl. Instrum. and Meth. {\bf A360} (1995) 481;\\
B.~Mours et~al., 
{\em The design, construction and performance of the ALEPH silicon vertex
detector}, 
Nucl. Instrum. and Meth. {\bf A379} (1996) 101.

\bibitem{priv:carlo}
C.~Ewerz, 
private communication.

\bibitem{Budnev:1974de}
V.M.~Budnev, I.F.~Ginzburg, G.V.~Meledin and V.G.~Serbo, 
{\em The two photon particle production mechanism, physical problems, 
applications, equivalent photon approximation}, 
Phys. Rept. {\bf 15} (1974) 181.

\bibitem{Nisius:1999cv}
R.~Nisius, 
{\em The photon structure from deep inelastic electron photon scattering}, 
Phys. Rept. {\bf 332} (2000) 165.

\bibitem{Brodsky:1997sd}
S.J.~Brodsky, F.~Hautmann and D.E.~Soper, 
{\em Virtual photon scattering at high energies as a probe of the short
distance pomeron}, 
Phys. Rev. {\bf D56} (1997) 6957.

\bibitem{Schuler:1997ex}
G.A.~Schuler, 
{\em Two-photon physics with GALUGA 2.0}, 
Comput. Phys. Commun. {\bf 108} (1998) 279.

\bibitem{l3_double:1999}
L3 Collaboration, 
{\em Measurement of the cross-section for the process 
$\gamma^* \gamma^* \rightarrow $ hadrons at LEP}, 
Phys. Lett. {\bf B453} (1999) 333.

\bibitem{opal_double:2000}
OPAL Collaboration, 
{\em Measurement of the hadronic cross-section for the scattering of two 
virtual photons at LEP}, 
Eur. Phys. J. {\bf C24} (2002) 17.

\end{thebibliography}
\end{document}